\documentclass[a4paper,11pt]{article}
\pdfoutput=1 

\usepackage{jheppub} 

\usepackage[T1]{fontenc}
\usepackage{xcolor,soul}
\usepackage{amsmath}
\usepackage{amssymb}
\usepackage{amscd}
\usepackage{stmaryrd}
\usepackage{bm}
\usepackage{courier}  
\usepackage{graphicx}

\newcommand{\beq}{\begin{equation}}
\newcommand{\eeq}{\end{equation}}
\newcommand{\ba}{\begin{array}}
\newcommand{\ea}{\end{array}}
\newcommand{\beqa}{\begin{eqnarray}}
\newcommand{\eeqa}{\end{eqnarray}}

\def\R{\mathbb R}

\title{\boldmath Color confinement at the boundary of the conformally compactified $\mathrm{AdS}_5$}

\author[a,c]{M. Kirchbach}
\author[c,e]{T. Popov,}
\author[b,d,1]{J. A. Vallejo\note{Corresponding Author.}}

\affiliation[a]{Instituto de F\'{i}sica, Universidad Aut\'onoma de
     San Luis Potos\'i,\\
     Av.\ Manuel Nava 6, Zona Universitaria, 78290 San Luis Potos\'i,
     \\
     M\'exico}
\affiliation[b]{Facultad de Ciencias, Universidad Aut\'onoma de
     San Luis Potos\'i,\\
     Av. Parque Chapultepec 1570, Lomas del Pedregal, 78290 San Luis Potos\'i,
     \\
     M\'exico}     
\affiliation[c]{Institute for Nuclear Research and Nuclear Energy,  Bulg.\ Acad.\ of Sc.
     \\
       Bul.\ Tsarigardsko Chauss\'ee 72, 1784  Sofia,\\
       Bulgaria}
\affiliation[d]{Departamento de Matem\'aticas Fundamentales, Facultad de Ciencias, UNED,\\
 C. Senda del Rey s/n, 28040 Madrid,\\
 Spain}
\affiliation[e]{American University in Bulgaria,\\
  8 Svoboda Bachvarova St., Blagoevgrad 2700,\\
 Bulgaria}
\emailAdd{mariana@ifisica.uaslp.mx} \emailAdd{tpopov@aubg.edu}
\emailAdd{jvallejo@mat.uned.es}

\abstract{The topology of closed manifolds forces interacting charges to appear in pairs. We take advantage of this property in the setting of the conformal boundary of $\mathrm{AdS}_5$ spacetime, topologically equivalent to the closed manifold $S^1\times S^3$, by considering the coupling of two massless opposite charges on it. Taking the interaction potential as the analog of Coulomb interaction (derived from a fundamental solution of the $S^3$ Laplace-Beltrami operator), a conformal $S^1\times S^3$ metric deformation is proposed, such that free motion on the deformed metric is equivalent to motion on the round metric in the presence of the interaction potential. We give explicit expressions for the generators of the conformal algebra in the representation induced by the metric deformation. 

By identifying the charge as the color degree of freedom in QCD, and the two charges system as a quark--anti-quark system, we argue that the associated conformal wave operator equation could  provide a realistic quantum mechanical description of the simplest QCD system, the mesons.

Finally, we discuss the possibility of employing the compactification radius, $R$, as another scale along $\Lambda_{QCD}$, by means of which, upon reparametrizing  $Q^2c^2$ as  $\left( Q^2c^2 +\hbar^2 c^2/R^2\right)$, a perturbative treatment of processes in the infrared could be approached.}

\begin{document} 
\maketitle
\flushbottom

\section{Introduction}\label{sec:intro}
 
On flat, Minkowskian $(d+1)-$spacetime $\mathbb{R}^{1,d}$ (including its null cone $\mathcal{C}^{1,n}$),
QED is an obvious candidate for a conformally invariant quantum field theory. It has the feature that propagation of a single
charge is possible. If one wants to force the appearance of charges in dipole configurations, a possibility is to consider a closed spacetime as explained in Section \ref{sec:chargeneu}, and one way to achieve this is through conformal compactification. As we will
see in Section \ref{sec:minkowski}, there is an intimate relation between the conformal compactification of Minkowski spacetime $\mathbb{R}^{1,d-1}$ and the boundary of the conformally compactified Anti-De Sitter spacetime $\mathrm{AdS}_{d+1}$, such that, for the particular case of $d=4$, we get in both cases the topology of the product of spheres $S^1\times S^3$.

A theory which describes (at least below certain pretty high energy scales) exclusively  charge-neutral constituent systems is QCD, the quantum field theory of strong interaction, in which the so far reliably reported strongly interacting particles, the hadrons, are all color neutral, a phenomenon referred to as color confinement. For such a theory, the consideration of the compactified Minkowski spacetime as an internal space, could provide an adequate explanation for the non-observability of free color-electric  charges. On the other hand, the consideration of QCD at the boundary of $\mathrm{AdS}_5$ is independently  supported by the AdS/CFT correspondence, according to which strongly coupled gravity in the bulk dually appears at the $\mathrm{AdS}_5$ boundary as a weakly coupled conformal field theory, conjectured to be QCD in the perturbative regime.

The simplest two-body system which any well-defined quantum field theory should allow, can be described at the quantum mechanical level by a wave equation with  a potential related to instantaneous one gauge-boson exchange, the Hydrogen atom being the paradigm for such an example. Indeed, the Coulomb potential, a harmonic solution to the flat-space Laplace equation, and  the spacelike image of the instantaneous one-photon exchange, implements the conformal symmetry of QED, and allows for a fairly precise description of the properties of this atom. Compared to  this, the quantum mechanical description of the simplest strong interacting systems, the mesons, presents itself as significantly unsatisfactory,  as it is  plagued by several deficiencies. The free parameters in the existing quark  potential models are numerous, and they are, as a rule, unrelated to the fundamental parameters of QCD.

It is the goal of the present work to show that, starting from the boundary of the conformally compactified $\mathrm{AdS}_5$, realized as the conformally compactified Minkowski spacetime,
the theory of strong interaction, QCD , is next to QED as a further example of a conformal quantum field theory which can be realistically represented by a quantum mechanical wave equation.

The counterpart to the Coulomb potential is found to be the cotangent function, constructed out of a fundamental solution of the Laplace-Beltrami operator on $S^3$ as a dipole potential. This reflects the already mentioned topological property of closed spaces: on them, no free charges are allowed. Thus, this class of spaces is apt to host charge neutral systems, such as color charge-neutral hadron dipoles. We also show that the strong interaction induced by such dipoles can be switched on and off by a properly chosen conformal deformation of the metric of the conformally compactified Minkowski spacetime.

The paper is structured as follows. In the next section, we briefly review the technique of conformal compactification, with emphasis on the relation between $\mathrm{AdS}_5$ and the Minkowski spacetimes. Section \ref{sec:chargeneu} contains the explanation on why closed spaces do not support free charges, and the modifications of the fundamental solutions of elliptic operators, such as the Laplacian, that this fact imposes.  In sections \ref{sec:coordinates} and
\ref{sec:waveops} we introduce a definition of the conformal wave operator on the conformally compactified Minkowski spacetime and present the parametrization of this space by hyper-spherical coordinates, summarizing the essential facts on conformal wave operators and their transformations under conformal deformations of the metric. Section \ref{sec:freepart} is devoted to the quantum motion of a free massless scalar object on the space thus defined. In section \ref{sec:perturbedpot} we consider the {\it ad hoc} extension of the conformal wave operator previously defined  by  a particular potential, constructed out of a fundamental solution of
the $S^3$ Laplacian. This case will sometimes  be referred to as the `interacting particle equation'.
Section \ref{sec:harmpot} deals with a critical issue, namely: we present a mechanism of inclusion of the interaction potential into the conformal wave operator, through a properly designed deformation of the conformally compactified Minkowski spacetime $\mathcal{M}^{1,3}$, a procedure which provides a further argument in favor of the conformal symmetry of the  proposed potential. 
According to the discussion given in sections \ref{sec:freepart} and \ref{sec:harmpot}, motion on $\mathcal{M}^{1,3}$ under this potential, is equivalent to free motion on the  conformally deformed metric. To the amount this equation is compatible with both the conformal symmetry of QCD in the infrared, signaled by the opening of the conformal window \cite{Deur} on the one side, and the color-electric charge confinement on the other, we suggest it as a candidate for the quantum mechanical limit of QCD.
Finally, in section \ref{sec:phenomenology} we summarize applications of the aforementioned equation to the physics of hadrons, illustrating its realistic predictive power.
In addition, we also discuss there the possibility of using the ${\mathcal M}^{1,3}$ compactification radius as a scale in QCD (next to
$\Lambda_{QCD}$) and reparametrize the negative square $-q^\mu q_\mu=Q^2$ of the transferred four momentum, $q^\mu$, by $\left(Q^2c^2 +\hbar^2c^2/R^2\right)$, a procedure which would lead to a walking of the strong coupling toward a fixed value at $Q^2=0$ and thus allowing for a regular QCD there. The paper ends with a brief conclusion and two appendices.
The first one lists the explicit expressions for the $so(2,4)$ generators of the isometry group $SO(2,4)$ of the
conformally compactified Minkowski spacetime, and leads to the second appendix, in which we determine  
the $so(2,4)$ generators in the representation induced by the conformal deformations of the metric on it.

\section{Minkowski spacetime and the boundary of $\mathrm{AdS}_5$}\label{sec:minkowski}

Conformal compactification is a convenient way of `bringing infinity to a finite distance' in 
spaces endowed with an indefinite scalar product $\left\langle\cdot ,\cdot\right\rangle$.
The key idea is to notice that conformal inversion
\[
x\mapsto \frac{1}{\left\langle x, x\right\rangle},
\]
sends points close to the null cone $\mathcal{C}=\{ x: \left\langle x, x\right\rangle=0\}$ to 
points `close to infinity', and vice versa. If we start with a Minkowskian spacetime 
$\mathbb{R}^{1,d}$ endowed with coordinates $(x_0,x_1,\ldots,x_d)$ and the scalar product
\[
\left\langle x, y\right\rangle =-x_0y_0+\sum^{d}_{i=1}x_iy_i\,,
\]
we can look for a bigger space that includes the points at infinity onto which the null cone of 
$\mathbb{R}^{1,d}$ will be mapped. To this end, consider the pseudo-Euclidean space 
$\mathbb{R}^{1+1,d+1}$ with coordinates $(x_{-1},x_0,x_1,\ldots,x_d,x_{d+1})$ and the scalar 
product
\[
\left\langle x, y\right\rangle =-x_{-1}y_{-1}-x_0y_0+\sum^{d+1}_{i=1}x_iy_i\,.
\]

Let $\mathcal{N}$ be the null cone of $\mathbb{R}^{1+1,d+1}$, that is,
\[
\mathcal{N}=\{x\in\mathbb{R}^{1+1,d+1}:\left\langle x, x\right\rangle=0  \}\,,
\]
and let $H$ be the intersection of $\mathcal{N}$ with the hyperplane $x_{-1}-x_{d+1}=1$, so
\[
H=\{x\in\mathbb{R}^{1+1,d+1}:-x^2_{-1}-x^2_0+ \sum^{d+1}_{i=1}x_iy_i=0,\, x_{-1}-x_{d+1}=1\}\,.
\]
The subset $H$ is a representation of $\mathbb{R}^{1,d}$ in the sense that  there is an 
isometry $\Phi:\mathbb{R}^{1,d}\to H\subset\mathbb{R}^{1+1,d+1}$, explicitly given by
\[
\Phi(x_0,x_1,\ldots,x_d)=\left(\frac{x^2+1}{2},x_0,x_1,\ldots,x_d,\frac{x^2-1}{2}\right)\,,
\]
where $x^2=-x^2_0+\sum^{d}_{i=1}x^2_i$. 

What are the `points at infinity' of $\mathbb{R}^{1,d}$ in this picture? Notice that
the generatrices of the null cone $\mathcal{N}$ intersect $H$ at most once, but there 
are some of them which do not cross $H$: those which are contained in hyperplanes parallel 
to $x_{-1}-x_{d+1}=0$. In other words, the generatrices of $\mathcal{N}$ that do not intersect 
$H=\Phi\left(\mathbb{R}^{1,d}\right)$ are those determined by null vectors
$(x_{-1},x_0,x_1,\ldots,x_d,x_{d+1})$ such that $x_{-1}=x_{d+1}$, and they coincide with 
the elements of $\mathcal{N}$
such that $(x_0,x_1,\ldots,x_d)$ lies on the null cone of $\mathbb{R}^{1,d}$. 
Hence, the `points at infinity' (corresponding
to the null cone of $\mathbb{R}^{1,d}$) are in correspondence with the generatrices of 
$\mathcal{N}\subset\mathbb{R}^{1+1,d+1}$
that do not intersect $H$.

In order to get a subspace of $\mathbb{R}^{1+1,d+1}$ containing both $H=\Phi\left(\mathbb{R}^{1,d}\right)$ 
and its points at infinity, we introduce the two radial coordinates in $\mathbb{R}^{1+1,d+1}$:
\begin{align*}
r^2 &:= x^2_{-1}+x^2_0,\\
s^2 &:= \sum^{d+1}_{i=1}x^2_i\,,
\end{align*}
and observe that the null cone $\mathcal{N}\subset\mathbb{R}^{1+1,d+1}$ is determined by the equation $r^2=s^2$, so its
intersection with the hyper-sphere $r^2+s^2=2$ is given by the equations
\[
r=1=s\,,
\]
topologically corresponding to a product of spheres $S^1\subset\mathbb{R}^{1+1},S^d\subset\mathbb{R}^{1+d}$. However,
each generatrix of the null cone $\mathcal{N}$ intersects the product $S^1\times S^d$ at two antipodal points, which
must be identified in order to get a unique correspondence with `points at infinity' for $\mathbb{R}^{1,d}$. Thus, the
space $\mathcal{M}^{1,d}$ defining the conformal compactification of $\mathbb{R}^{1,d}$, is given by the subspace of
$\mathbb{R}^{1+1,d+1}$ determined by the intersection of null cone $\mathcal{N}\subset \mathbb{R}^{1+1,d+1}$ (with
equations $r^2=s^2$) and the hyper-sphere $r^2+s^2=2$, where antipodal points are identified, resulting in
\[
\mathcal{M}^{1,d}=(S^1\times S^d)/\mathbb{Z}_2\,.
\]

We are interested in the topology of this space. To this end, notice that we have a bundle
\[
S^d\hookrightarrow (S^1\times S^d)/\mathbb{Z}_2\to S^1/\mathbb{Z}_2\simeq S^1\,.
\]
When $d$ is odd, the antipodal map on the sphere $S^d$ is orientation-preserving. It is well known that if $M$ is a connected,
oriented manifold, and $G$ is a discrete group acting freely and properly on $M$, then $M/G$ is orientable if and only if
$G$ is orientation preserving. Therefore, the quotient $(S^1\times S^2)/\mathbb{Z}_2$ is orientable when $d$ is odd (in what
follows, we will restrict ourselves the case $d=3$). It being the total space of the bundle above, the classification
theorem of bundles over $S^1$ \cite{Steenrod} guarantees that this must be a product bundle. Summing it up, we get (as in \citep{PR,Scho})
\[
\mathcal{M}^{1,d}=(S^1\times S^d)/\mathbb{Z}_2\simeq S^1\times S^d\mbox{ }(d\mbox{ }odd)\,,
\]
which, as a product of closed spaces, is closed.

Let us now consider Anti-De Sitter spacetime, $\mathrm{AdS}_{d+2}$, which is the one-sheet hyperboloid into the Euclidean
space $\mathbb{R}^{2,d}$, endowed with coordinates $(y_{-1},y_0,y_1,\ldots,y_d)$ and a pseudo-Euclidean metric of
signature $(1,1,-1,\ldots,-1)$, given by the equation
\begin{equation}\label{eqhypads}
y^2_{-1}+y^2_0-\sum^{d}_{i=1}y^2_i=L^2\,.
\end{equation}
Notice that the isometry group of $\mathrm{AdS}_{d+2}$, as determined by the quadric \eqref{eqhypads}, is $SO(2,d)$. The
induced metric $\tilde{g}$, on $\mathrm{AdS}_{d+2}$ can be computed from the following, explicit expression for the embedding
$\mathrm{AdS}_{d+2}\hookrightarrow\mathbb{R}^{2,d}$:
\begin{align*}
y_{-1} &=L\cosh\rho\sin\tau, \\
y_{0} &=L\cosh\rho\cos\tau, \\
y_i &=L\sinh\rho\,\Omega_i\,,
\end{align*}
where $\Omega$ collectively denotes the spherical coordinates on $\sum^n_{i=1}\Omega^2_i=1$. It is readily found that
\[
\tilde{g}=L^2(\cosh^2\rho\mathrm{d}\tau^2-\mathrm{d}\rho^2-\sinh^2\rho\mathrm{d}\Omega^2_{d-1})\,.
\]
Also notice that $\tilde{g}$ is a Lorentzian metric, with $\tau$ a periodic coordinate $\tau\in S^1$, and $\rho$ a radial
one, $\rho\in\, [0,+\infty [\,$. It is common `to bring infinity to a finite distance' by making a further change of coordinates
from $\rho$ to $\theta\in\, [0,\pi/2[\,$ through
\[
\sinh\rho =\tan\theta\,,
\]
thus obtaining the metric of the so-called conformally compactified $\mathrm{AdS}_{d+2}$, a spacetime topologically equivalent to $S^1\times S^3\times [0,\pi/2[$ with 
metric\footnote{The metric given in \eqref{eqmetads} does not define a causal spacetime, due to the periodic character of $\tau$. For this reason, sometimes the covering space $\mathbb{R}$ of $S^1$ is considered instead, but this is an issue
that will not concern us in this paper.} $g$ given by
\begin{equation}\label{eqmetads}
g=\frac{L^2}{\cos^2\theta}\left( \mathrm{d}\tau^2-\mathrm{d}\theta^2-\sin^2\theta\mathrm{d}\Omega^2_{d-1} \right)\,.
\end{equation}

The underlying space has an asymptotic boundary at infinity that can be computed (using invariance under Weyl transformations) by the rule
\[
\partial (X\times Y)=(\partial X\times \overline{Y})\cup (\overline{X}\times\partial Y)\,,
\]
where $\partial$ is the boundary operator and $\overline{X}$ denotes the topological closure of the corresponding space, resulting in the topological space $S^1\times S^3$. Thus, the boundary of the conformally compactified $\mathrm{AdS}_{d+2}$ is topologically equivalent to the conformal compactification of Minkowski
spacetime $\mathbb{R}^{1,d}$.

 Introducing the notation,
 ${\vec y}\, ^2= y^2_1+y^2_2 +y^2_3 +y^2_4 \,$, the equation for the Anti-De Sitter hyperboloid can be rewritten as
 \begin{eqnarray}
{\vec y}\, ^2+L^2=y^2_{-1}+y^2_0\, .
\label{frnsd}
 \end{eqnarray}
 The boundary at infinity corresponds to 
$|{\vec y}\, |\to \infty$, taken at a fixed point of $S^1\times S^3$, and 
can be asymptotically identified with the null-ray cone, $L^2=0$, to the amount any finite 
value of $L$ becomes negligible compared to the above mentioned limit \cite{Frondsdal}.

Within the context of  the $AdS_5/CFT_4$ gauge-gravity duality conjecture, on the $\mathrm{AdS}_5$ boundary (identified as above with the null-ray cone of ${\mathbb R}^{2,4}$ \cite{Frondsdal}) conformal field theories in their perturbative regimes can reside, some of which   could   appear dual to strongly coupled gravity in the bulk. One theory of this kind is Quantum Electrodynamics (QED), which  describes propagation of single charges. As  a process which is only possible in open spaces (see section \ref{sec:chargeneu}), it necessarily resides in ${\mathbb R}^{1,3}$ and its corresponding  light cone, denoted 
${\mathcal C}^{1,3}$ \cite{Frondsdal}. At the quantum mechanical level it is implemented by the Hydrogen atom, described by the Coulomb potential, the Fourier transform to position space of the instantaneous photon propagator, and a harmonic solution to the flat space Laplace equation.
Next to QED, another theory of this kind, the gauge theory of strong interaction called Quantum Chromodynamics (QCD), can be considered. In its regime of asymptotic freedom, the conformally  invariant ultraviolet, it also can reside on ${\mathcal C}^{1,3}$.
In its infrared regime the  viable degrees of freedom systems are those of confined color-electric charges,  the strong coupling walks in the limit of vanishing momentum transfer toward a fixed value \cite{Deur}, thus opening the so called conformal window, and possibly an avenue toward a  perturbative treatment in
the infrared. All of these are reasons for which QCD can be placed on ${\mathcal M}^{1,3}$. In parallel to QED, one can expect QCD in this regime to also have a well behaved quantum mechanical limit, and allow for a description of the simplest
quark--anti-quark  two-body  systems  by a wave equation with a potential, taken to be a harmonic function to the $S^3$ Laplacian.
One of the goals of the present work is to advocate for the construction of such an equation.

So far in the literature, QCD has been treated at both 
${\mathcal C}^{1,3}$  and ${\mathcal M}^{1,3}$ geometries. On ${\mathcal C}^{1,3}$, of relevance to the  perturbative ultraviolet regime of asymptotic  freedom,  the light front QCD has been developed \cite{BrTrm}, while in \cite{Hands}
${\mathcal M}^{1,3}$ has been used in a perturbative approach to the QCD
phase transition, in which the scale necessary for the introduction of temperature  has been provided by the hyper-radius of ${\mathcal M}^{1,3}$, treated as a small hyper-spherical box with chemical potential.

\section{Charge neutrality on closed spaces}\label{sec:chargeneu}

We have mentioned that, ultimately, we are 
going to modify the free Hamiltonian on $S^3$ by adding a potential containing the 
cotangent function. In this section, we will try to understand the geometric origin of this potential, 
and for this, we need to discuss the fundamental solutions of the Laplacian operator.

Consider for a moment an open subset  $U\subset\mathbb{R}^n$, with $n\geq 2$, 
and the Laplacian $\Delta$  on functions 
constructed from the Euclidean metric on the ambient space. 
A fundamental solution of $\Delta$ centered on $\mathbf{y}\in U$,
is a distributional solution $G$ of the equation
\[
\Delta G=\delta_\mathbf{y}\,,
\]
where $\delta_\mathbf{y}$ is the Dirac singular distribution, acting on test functions 
$\varphi\in\mathcal{C}^\infty_0(U)$ as
\[
\langle\delta_\mathbf{y};\varphi\rangle =\varphi(\mathbf{y})\,.
\]

It is well known that, defining $f:\,]0,\infty[\,\to\,]0,\infty[$ by
\[
f(r)=\begin{cases}
\dfrac{1}{n(2-n)w_{n-1}}|r|^{2-n}\mbox{ if }n\neq 2,\\[8pt]
\dfrac{1}{2\pi}\ln r\mbox{ if }n=2\,,
\end{cases}
\]
where $w_{n-1}$ is the volume\footnote{Just for reference, its value is
\[
w_{n-1}=\frac{2\pi^{n/2}}{\Gamma(n/2)}\,.
\]} of the unit sphere $S^{n-1}\subset\mathbb{R}^n$, then the function
$G:\mathbb{R}^n\times\mathbb{R}^n\to\mathbb{R}$ given by
\[
G(\mathbf{x},\mathbf{y})=f(\|\mathbf{x}-\mathbf{y}\|),
\]
is such that, for each fixed $\mathbf{y}\in\mathbb{R}^n$, the new function
\[
G_{\mathbf{y}}(\mathbf{x})=G(\mathbf{x},\mathbf{y})=f(\|\mathbf{x}-\mathbf{y}\|),
\]
satisfies (among other properties):
\begin{enumerate}
\item It is harmonic on $\mathbb{R}^n-\{\mathbf{y}\}$, that is,
\[
\Delta_\mathbf{x}G_\mathbf{y}(\mathbf{x})=0\,.
\]
\item $G_\mathbf{y}$ is actually a distribution and the equation
\begin{equation}\label{eq26}
\Delta G_\mathbf{y}=\delta_\mathbf{y},
\end{equation}
holds in the distributional sense.
\end{enumerate}

In other words, the $G_\mathbf{y}$ are fundamental solutions of the (Euclidean) 
Laplacian on open subsets of $\mathbb{R}^n$.
Notice that fundamental solutions are never unique, as adding up a harmonic function 
to any one of them, we obtain a new fundamental solution.

Of course, the meaning of \eqref{eq26} is that, for any test function 
$\phi\in\mathcal{C}^\infty_0(U)$, we have
\begin{equation}\label{eq27}
\langle\Delta G_\mathbf{y},\phi\rangle = \langle G_\mathbf{y},\Delta\phi\rangle =
\langle\delta_\mathbf{y},\phi\rangle 
=\phi(\mathbf{y})\,,
\end{equation}
that is, equating the second and fourth terms:
\[
\int_U G(\mathbf{x},\mathbf{y})\Delta\phi(\mathbf{x})=
\int_U G_\mathbf{y}(\mathbf{x})\Delta\phi(\mathbf{x})=\phi(\mathbf{y})\,,
\]
so, in a sense, the fundamental solution is an inverse of the Laplacian.

If we now have a manifold $M$ instead of just an open subset of an Euclidean space, 
in order to make sense of integral formulas such as the one above, we need to impose 
some topological restrictions on $M$. For instance, requiring $M$ to be compact we 
guarantee that it admits Riemannian metrics and then, we can construct both the
Laplacian on $\mathcal{C}^\infty(M)$ and the measure on $M$ determined by the 
Riemannian volume form, so the individual terms appearing in the preceding equations
make sense. Unfortunately, however, the equations themselves do not.

To understand why this is so, consider a compact manifold $M$ instead of $U\subset\mathbb{R}^n$, and take a point
$y\in M$. Suppose that \eqref{eq26} or its equivalent \eqref{eq27} were true. We can then consider the function
$\phi\in\mathcal{C}^\infty_0(M)=\mathcal{C}^\infty(M)$ as being represented in the form $y\mapsto \langle\delta_y,\phi\rangle$. Then, for any $\varphi\in\mathcal{C}^\infty(M)$:
\[
\int_M\phi(y)\varphi(y)=\langle\langle\delta_y,\phi\rangle;\varphi(y)\rangle
=\langle\langle\Delta_xG_y(x);\phi(x)\rangle ;\varphi(y)\rangle\,.
\]

But, from the definition of distributional derivatives,
\[
\langle\langle\Delta_xG_y(x);\phi(x)\rangle ;\varphi(y)\rangle =
\langle\langle G_y(x);\Delta_x\phi(x)\rangle ;\varphi(y)\rangle\,,
\]
and taking as a particular case the constant function $\phi=\mathbf{1}_M\in\mathcal{C}^\infty(M)$, we get
$\Delta_x\phi(x)=0$, leading to the identity
\[
\int_M \varphi(y)=0,
\]
for any $\varphi\in\mathcal{C}^\infty(M)$, which (again by considering constant functions) is absurd.

Notice that this problem is absent in open subsets of $\mathbb{R}^n$, as constant functions do not have compact support
there and can not be taken as test functions. It is possible to compensate, in a compact manifold, for the presence of 
constant test functions if we modify the definition of fundamental solution, so that it now reads
\[
\Delta G_y=\delta_y +c\,,
\]
with $c\in\mathbb{R}$ an appropriate constant. In order to determine it, let us repeat the preceding calculations,
this time arriving at
\[
\langle\langle G_y(x);\Delta_x\phi(x)\rangle ;\varphi(y)\rangle =
\langle\langle\delta_y+c,\phi(x)\rangle ;\varphi(y)\rangle =
\langle\phi(y);\varphi(y)\rangle + \left\langle c\int_M\phi(x) ;\varphi(y)\right\rangle\,,
\]
that is, taking $\phi=\mathbf{1}_M\in\mathcal{C}^\infty(M)$ as before,
\[
0=\int_M\varphi(y) +c\left(\int_M\mathbf{1}\right)\left(\int_M\varphi(y)\right)\,,
\]
which is consistent if we take \cite{aubin}
\[
c=-\frac{1}{\mathrm{vol}(M)}\,.
\]

We recapitulate these observations in the following definition. Let $M$ be a compact manifold with finite volume
$\mathrm{vol}(M)$. Take any Riemannian metric $g$ on $M$ and let $\Delta$ be its associated Laplacian. Then, a
fundamental solution of $\Delta$ is any function $G:M\times M\to \mathbb{R}$ satisfying, for any $y\in M$ fixed,
\begin{equation}\label{eq28}
\Delta_xG(x,y)=\delta_y-\frac{1}{\mathrm{vol}(M)}\,.
\end{equation}

Given a fundamental solution $G$, if $u\in\mathcal{C}^\infty(M)$ is a harmonic function (that is, $\Delta u=0$), then
$G+u$ is another fundamental solution. However, on a compact manifold, Liouville's theorem implies that the harmonic
functions are precisely the constants. Thus, given a particular $G$, any other fundamental solution will have the form
$G+c$, where $c$ is a constant. 

For the case of the three sphere $S^3$, parametrized by hyper-spherical coordinates
$(\rho,\theta,\varphi)$, we have  $\mathrm{vol}(M)=2\pi^2$, and a direct computation
\cite{Brigita} shows that a fundamental solution with center $x$ given at the pole 
$\rho =0$ is precisely given by the cotangent-like function as
\begin{equation}\label{cotsolfund}
G_0(\rho,\theta,\varphi)=\frac{1}{4\pi^2}(\pi-\rho) \cot \rho\,.
\end{equation}
Physically, the relevance of this result is the following. The equation
\[
\Delta V(x,y)=Q\delta_y,
\]
describes the Coulomb potential created at $x$ by a single charge $Q$ placed at $y$. Thus, we conclude that on a sphere (and, more generally, on any closed space) a Coulomb-type interaction does not admit isolated charges. One can wonder what will be the minimal charge configuration possible on such spaces. In the particular case of a sphere, its large symmetry group
allows for another definition of fundamental solution for the Laplacian, taking a base point and its antipodal at the same time, leading to the equation \cite{cohl}
\begin{equation}\label{cohl}
\Delta G(x,x')=\delta (x,x')-\delta (-x,x')\,,
\end{equation}
which is nothing but the equation for the well known dipole Green function giving rise to a cotangent potential. We insist in that this approach is not generally applicable. However, it can be proved that both approaches,
as given by \eqref{eq28} and \eqref{cohl}, coincide on spheres \cite{chap}. Hence, closed spaces can support only charge-neutral systems, the dipole being the most elementary one, for
which a cotangent-like function as in \eqref{cotsolfund} is the equivalent of the Coulomb potential.

\section{Coordinates and conformal wave  operators on the compactified Minkowski spacetime}\label{sec:coordinates}

We begin by  introducing polar coordinates on the $\mathrm{AdS}_5$ null-ray cone  according to
 \begin{eqnarray}
y^{-1}&=&R\sin\alpha,\nonumber\\
y^0&=&R\cos\alpha,\nonumber\\
y^1&=&R\sin\chi\sin\theta\cos\varphi ,\nonumber\\
y^2&=&R\sin\chi\sin\theta\sin\varphi , \nonumber\\
y^3&=&R\sin\chi\cos\theta,\nonumber\\
y^4&=&R\cos\chi, \quad \chi,\theta\in [0,\pi], \quad \alpha, \varphi\in [0,2\pi[\,,
\label{polar_cmpct}
 \end{eqnarray}
where $\alpha$ is the so-called circular conformal time. In these variables the metric of ${\mathcal M}^{1,3}\simeq S^1\times S^3$, reads \citep{Huguet,BirrelDavis}
 \begin{equation}
 g=d\alpha^2 -\left(
   d\chi^2 +\sin\chi^2( d\theta^2 +\sin^2\theta d\varphi^2)\right)\,.
\label{Ads52}
 \end{equation}

A massless field conformally coupled to this metric is described by means of the equation,
 \begin{equation}
\left[\Box_{S^1\times S^3} (\alpha,\chi,\theta,\varphi) +1 \right]\Psi(\alpha,\chi,\theta,\varphi)=0\,, \label{CCMSF_1}
 \end{equation}
Here, $\Box_{S^1\times S^3} (\alpha, \chi,\theta,\varphi)$ stands for the
 conformal wave  operator on ${\mathcal M}^{1,3}$,
 \begin{equation}
 \Box _{S^1\times S^3}(\alpha,\chi,\theta,\varphi) =\frac{\partial^2}{\partial\alpha^2} -\Delta_{S^3}(\chi,\theta,\varphi)\,,
\label{C2} 
 \end{equation} 
  $\Delta_{S^3}(\chi,\theta,\varphi)$ is the Laplace-Beltrami operator on the three dimensional hyper-sphere,  $S^3$,  
\begin{equation}
-\Delta_{S^3}(\chi,\theta,\varphi)={\mathcal K}(\chi,\theta,\varphi)=-\frac{1}{\sin^2\chi}\frac{\partial }{\partial \chi }\sin^2\chi \frac{\partial }{\partial \chi} +\frac{{\bf L}^2(\theta,\varphi)}{\sin^2\chi}\,, \label{C3}
\end{equation} 
while  ${\bf L}^2(\theta,\varphi)$ and ${\mathcal K}(\chi,\theta,\varphi)$
denote the operators of the squared angular momenta in three and four-dimensional Euclidean 
spaces, respectively. Also, the unit radius, $R=1$, has been considered for simplicity.

Upon deformation of the above metric according to,
 \begin{eqnarray}
   \Omega^2(\alpha,\chi)ds^2&=&d{\tilde s}^2,
   \label{Hug_def}
\end{eqnarray}
with the conformal factor being chosen as
\begin{eqnarray}
  \Omega^2(\alpha,\chi)=\frac{2}{(1-\xi )\cos\chi +(1+\xi)\cos\alpha }
, &&  \quad n=4,
\label{Ads53}
 \end{eqnarray}
Minkowski, $dS_4$ and $AdS_4$ spacetimes can appear in depending on the $\xi$ value \cite{Huguet}. The conformal wave operators on these deformed metrics  are then obtained as,
 \begin{eqnarray}
   \Box_\xi(\alpha,\chi,\theta,\varphi)&=&\frac{1}{\Omega^2(\alpha,\chi)}
   \Box_{S^1\times S^3}(\alpha,\chi,\theta,\varphi)\nonumber\\
  & +&\frac{1}{\Omega (\alpha,\chi) }
   \left[(1+\xi)\sin\alpha \frac{\partial}{\partial_\alpha} -(1-\xi)\sin\chi
     \frac{\partial}{\partial_\chi}\right].
   \label{C1}
   \end{eqnarray}
 Correspondingly, the wave equation,
\begin{equation}
  \left[\Box_{\xi} (\alpha,\chi,\theta,\varphi) +2\xi\right]
  \Psi_\xi(\alpha,\chi,\theta,\varphi)=0,
\label{CCMSF}
\end{equation}
describes a massless field,
$\Psi_\xi(\alpha,\chi,\theta,\varphi)$, conformally coupled to the metric in (\ref{Hug_def}).

 \noindent
 
 Alternatively, one may also look at the Laplace operator, $\Delta_{{\mathbf R}^4}$ in ${\mathbf R}^4$,
 as expressed by means of  the hyper-spherical coordinates in (\ref{polar_cmpct}),
 \begin{eqnarray}
   \Delta_{{\mathbf R}^4}(R,\chi,\theta,\varphi)&=&
   \frac{1}{R^3}\frac{\partial }{\partial R} R^3 \frac{\partial}{\partial R}
   +\frac{1}{R^2}\Delta_{S^3 }(\chi,\theta,\varphi),
     \label{we1}
 \end{eqnarray}
 and allowing $R$ to grow in the fifth dimension according to $R=e^\tau$
 (sometimes $\tau$ is called the `conformal time'). This amounts to rewriting the
 conformal wave operator on $S^1\times S^3$ as 
   \begin{eqnarray}
     \Box_{S^1\times S^3} (\tau,\chi,\theta,\varphi)&=& e^{-2\tau}
     \left(
              \frac{\partial^2}{\partial \tau^2} +
                    \left(
                             {\mathcal K}(\chi, \theta,\varphi)+1
      \right)\right),
\label{we3}
   \end{eqnarray}
   with ${\mathcal K}(\chi,\theta,\varphi)$ defined in (\ref{C3}), 
which is again a conformal wave operator. Notice that, aside a conformal factor, it is
the operator defined in \eqref{CCMSF_1}.

Later on, we will show how, through a properly chosen conformal metric deformation, the  conformal wave operator $\Box_{S^1\times S^3} (\tau ,\chi,\theta,\varphi)$ in (\ref{we3}) can give a realistic quantum mechanical description of the color neutrality of hadrons in terms of a potential modeled on \eqref{cotsolfund}, that is, a fundamental solution of the Laplace-Beltrami operator on $S^3$.

\section{Generalities on conformal wave operators}\label{sec:waveops} 
In what follows, the notion of  conformal re-scaling of the metric and the associated conformal wave  operator  will be relevant, so we present here a brief resume of their properties \cite{Canzani}. The conformal wave operator $\Box_g$ associated to the metric $g$ is defined as
\begin{equation}
  \Box_g u=\left[
    \frac{\partial ^2}{\partial \tau^2}- \Delta_g +\frac{n-2}{4(n-1)}
    \mbox{Scal} _g \right] u\,,
\label{Canz1}
\end{equation}
where $\Delta_g  $ is the Laplace-Beltrami operator on the spacelike part of the
manifold with the $g$ metric, and Scal$_g$ is the scalar curvature. Under a conformal deformation,
\begin{equation}
g^*=e^{-2f}g\,,
\label{mtrc_dfrm}
\end{equation}
the $\Box_g$ operator transforms as,
\begin{equation}
\Box_{g^\ast}=e^{\frac{n+2}{2}f} \circ  \Box_g  \circ e^{-\frac{n-2}{2}f}\,.
\label{rsclng}
\end{equation}

Dragging the $e^{-\frac{n-2}{2}f}$ exponential from the very right to the very left, the equation (\ref{rsclng}) simplifies to
\begin{equation}
  \Box_{g^*} (u)= e^{2f}\left[
    \frac{\partial ^2}{\partial \tau^2}- \Delta_g +
(n-2)g^{ij}\frac{\partial f}{\partial x_j}\frac{\partial }{\partial x_i}
+  \frac{n-2}{4(n-1)}
    \mbox{Scal} _{g^*} \right] u\,,
\label{Canz2}
\end{equation}
where Scal$_{g^\ast}$ stands for the scalar curvature on the re-scaled metric, explicitly
\begin{eqnarray}
  \mbox{Scal}_{g^\ast}&=& \mbox{Scal}_g +\frac{4(n-1)}{n-2} e^{\frac{n-2}{2}f}\Box_ge^{-\frac{n-2}{2}f}.
  \label{Scalar_crv}
\end{eqnarray}
We here are interested in functions $u_{(g)}$ and $u_{(g^\ast )}$   for which  $\Box_g(u_{(g)})$ and
$\Box_{g^\ast}(u_{(g^\ast)})$ both vanish. For this particular case,
a basic result holds \cite{BirrelDavis}, which states that,
\begin{equation}
  \Box_g(u_{(g)})=0 \quad \mbox{if\,\, and\,\, only\,\, if}\quad \Box_{(g^\ast=e^{-2f}g)}\left(
  e^{\frac{n-2}{2}f}u_{(g)}\right) =0.
\label{fund_thrm}
\end{equation}
Indeed, replacing  $u_{(g)}$ in (\ref{rsclng})  by $u_{(g^\ast)}=e^{\frac{n-2}{f} f}u_{(g)}$, necessarily leads to,
\begin{equation}
\Box_{(g^\ast=e^{-2f}g)}(e^{\frac{n-2}{2}f}u_{(g^\ast)})=e^{\frac{n+2}{2}f} \Box_g u_{(g)}=0\quad \mbox{for}\quad  \Box_g u_{(g)}=0.
\label{rsclng_fndt}
\end{equation}
With the aid of (\ref{Scalar_crv}) and (\ref{Canz2}), a new equation can be
formulated  as,  
\begin{eqnarray}
&&  \left[\Box_g + {\mathcal V}_g \right]u_{(g^\ast)}=0,\nonumber\\
&&  {\mathcal V}_g= (n-2)g^{ij}\frac{\partial f}{\partial x_j}\frac{\partial }{\partial x_i}
+   e^{\frac{n-2}{2}f}\Box_ge^{-\frac{n-2}{2}f},
\label{Canz23}
\end{eqnarray}
and if solvable,  be interpreted as motion on the $g$ metric perturbed by the potential ${\mathcal V}_g$. In this fashion, potentials can be produced by metric deformations, an observation of interest in what follows. Notice that, in general, such potentials contain gradient terms.

The following two sections will be devoted to the study of two particular cases of these equations:
the free motion on $S^1\times S^3$, and the motion in a potential determined by a cotangent function.

\section{Free massless scalar particle on the compactified Minkowski space time}\label{sec:freepart}

According to what we have seen, free motion of a massless scalar particle on  $S^1\times S^3$ is described by means of $\Box_{S^1\times S^3}$ in \eqref{we3} as,

\begin{eqnarray}
&&\Box_{S^1\times S^3}(\tau,\chi,\theta,\varphi) \Psi(\tau,\chi,\theta,\varphi)=0,\nonumber\\
&&\Psi(\tau,\chi,\theta,\varphi)= e^{i\tau (K+1)} Y_{K\ell m}(\chi,\theta,\varphi),
\label{dAl1}
\end{eqnarray}
where $Y_{K\ell m}(\chi, \theta,\varphi)$  denote the eigenfunction of the
${\mathcal K}(\chi,\theta,\varphi)$ operator in (\ref{C3}), 
\begin{equation}
  {\mathcal K}(\chi,\theta,\varphi) Y_{K\ell m}(\chi,\theta,\varphi)=
  K(K+2)Y_{K\ell m}(\chi,\theta,\varphi),
\label{4Dam}
\end{equation}
with $K$ denoting the four-dimensional angular momentum value.
The ${\mathcal K}(\chi,\theta,\varphi)$  eigenfunctions are the well known four-dimensional  ultra-spherical harmonics,
\begin{eqnarray}
  Y_{K\ell m}(\chi,\theta,\varphi)&=&{\mathcal N}_{K\ell}S_{K\ell}(\chi)Y_{\ell m}(\theta,\varphi),
  \label{YKlm}\\
  S_{K\ell}(\chi)&=&\sin^\ell\chi G_{n}^{\ell +1}(\cos\chi), \quad K=n+\ell.
  \label{qsrad}
\end{eqnarray}
Here, $Y_{\ell m}(\theta,\varphi)$ are the ordinary spherical harmonics which  diagonalize ${\mathbf L}^2(\theta,\varphi)$,  $G_n^\nu(x)$ denote the Gegenbauer (ultra-spherical) polynomials, $S_{K\ell}(\chi)$ are `quasi-radial' functions, and ${\mathcal N}_{K\ell}$ denote normalization constants.

For future purposes it is convenient to remark that ${\mathcal K}(\chi,\theta,\varphi)$ in (\ref{C3}), acts in the same way on $Y_{K\ell m}(\chi,\theta,\varphi)$ as on \emph{any} linear superposition of the type 
$\sum_{\ell^\prime =0}^{K}a_{K\ell}^{\ell^\prime} Y_{K\ell^\prime m}(\chi,\theta,\varphi)$,
i.e. one has,
\begin{eqnarray}
  \Box_{S^1\times S^3}(\tau,\chi,\theta,\varphi)&& \left[e^{i(K+1)\tau}Y_{K\ell m}(\chi,\theta,\varphi)\right]=0=
    \nonumber\\
 \Box_{S^1\times S^3}(\tau,\chi,\theta,\varphi)&& \left[e^{i(K+1)\tau} \sum_{\ell^\prime =0}^{K}
  a_{K\ell}^{\ell^\prime} Y_{K\ell^\prime m}(\chi,\theta,\varphi)\right].\quad
\label{dAl1_1}
\end{eqnarray}
(also see \eqref{bss_eqv} below). Depending on the problem under consideration, one may choose to replace
in \eqref{4Dam} the action of 
${\mathbf L}^2(\theta,\varphi)$ on $Y_\ell ^m(\theta, \varphi)$ by its eigenvalue, $\ell (\ell+1)$, thereby reducing this equation to an equation in 
one variable, defining ${\mathcal K}(\chi)$ as,
\begin{eqnarray}
  {\mathcal K} (\chi)
  &=&
  \left[-\frac{1}{\sin^2\chi}\frac{\partial }{\partial \chi }\sin^2\chi \frac{\partial }{\partial \chi} +\frac{\ell (\ell +1)}{\sin^2\chi}\right]\,.
\label{point_1}
\end{eqnarray}

For the case of the linear superposition in the right-hand side in (\ref{dAl1_1}), it is clear
that we can replace $\ell^\prime (\ell^\prime +1)$ by $\ell (\ell +1)$, leading to,
\begin{equation}
{\mathcal K}(\chi)S_{K\ell}(\chi)=K(K+2)S_{K\ell}(\chi)\,.
\label{auxl_1}
\end{equation}

In fact, upon separating variables in the $e^{i(K+1)\tau}Y_{K\ell m}(\chi,\theta,\varphi)$ basis, equation (\ref{dAl1}) becomes the following {\it reduced} two-variable equation, 
\begin{eqnarray}
  &&\Box_{S^1\times S^3}(\tau,\chi) e^{i(K+1)\tau}S_{K\ell }(\chi)\nonumber\\
  &=& \left( \frac{\partial ^2}{\partial \tau^2}
  -\frac{1}{\sin^2\chi}\frac{\partial}{\partial\chi} \sin^2\chi \frac{\partial}{\partial \chi} +\frac{\ell (\ell +1)}{\sin^2\chi} +1 \right)e^{i(K+1)\tau}S_{K\ell }(\chi)\nonumber\\
&=&\left( \frac{\partial ^2}{\partial \tau^2}
  -\frac{\partial^2}{\partial\chi^2} -2\cot\chi \frac{\partial}{\partial \chi}+\frac{\ell (\ell +1)}{\sin^2\chi} +1\right)e^{i(K+1)\tau}S_{K\ell }(\chi).
  \label{instr1}
    \end{eqnarray}
Changing now the wave function as,
\begin{equation}
S_{K\ell}(\chi)\longrightarrow \frac{ \psi_{K\ell} (\chi)}{\sin\chi},
\label{var_chng}
\end{equation}
allows one to reach the following one-variable Schr\"odinger equation,
\begin{eqnarray}
  \left[
    -\frac{
      {\mathrm d}^2}
    {{\mathrm d}\chi^2} +\frac{\ell(\ell+1)}{\sin^2\chi} +1\right]
  \psi_{K\ell}  (\chi)&=&
  (E^2_{(K+1)}+1)\psi_{K\ell}(\chi),\nonumber\\\
E^2_{(K+1)} &=&(K+1)^2,
\label{SchrScrf}
\end{eqnarray}
which we shall consider as an equivalent description of the free motion on $S^3$ by means of a Schr\"odinger operator. Notice that the ${\mathcal K}(\chi)-$ and the
Schr\"odinger operator eigenvalues (spectra) differ by one unit, giving
$(K(K+2)+1)$ in (\ref{4Dam}) versus $\left[(K+1)^2+1\right]$ in (\ref{SchrScrf}), respectively.

\section{Massless scalar  particle on $S^1\times S^3$ in the presence of an  {\it ad hoc} conformal potential}\label{sec:perturbedpot}

\vspace{0.31cm}
Now the question may be posed regarding the motion on ${\mathcal M}^{1,3}\simeq S^1\times S^3$ of a scalar particle within a dipole potential obtained from a fundamental solution of the Laplace-Beltrami operator on $S^3$.
This potential is well known \cite{Brigita} and is given (modulo an additive constant) by multiples of the cotangent function (c.f. (\ref{Al_Bel} below). Then, the {\it ad hoc} extension of \eqref{instr1} by  such potential, leads to following wave equation:

\begin{eqnarray}
  \left[ \Box_{S^1\times S^3}(\tau,\chi,\theta,\varphi) \pm 2b\cot\chi +\frac{\alpha_K^2}{4} \right]\psi(\tau,\chi,\theta,\varphi)&=&0,\nonumber\\
  \psi(\tau,\chi,\theta,\varphi)= e^{i\tau (K+1)}
  \Phi_{K\ell m}(\chi,\theta,\varphi).&&
\label{Frnds3}
\end{eqnarray}

By separating variables, the spatial part of the preceding equation gives rise to the following eigenvalue problem,
\begin{eqnarray}
  \left[ -\sin^2\chi \frac{\partial}{\partial \chi}\sin^2\chi \frac{\partial}{\partial \chi} +1+ V_{tRM}(\chi)\right]\Phi_{K\ell m}(\chi,\theta,\varphi)&=&\nonumber\\
  (K+1)^2 \Phi_{K\ell m}(\chi,\theta,\varphi),&&\nonumber\\
V_{tRM}(\chi) =\frac{\ell (\ell +1)}{\sin^2\chi} \pm \alpha_K(K+1) \cot\chi &+&
\frac{\alpha_K^2}{4 }.
\label{Frnds8}
\end{eqnarray}

The potential $V_{tRM}(\chi)$ is the trigonometric Rosen-Morse potential, and the perturbation it causes on free motion is well studied; in particular, the solutions to \eqref{Frnds8}  are known to be of the form
\begin{equation}
 \Phi_{K\ell m}(\chi,\theta,\varphi)= F_{K\ell}(\chi) Y_\ell^m(\theta,\varphi)\,,\label{sol_CCl}
\end{equation}
where
\begin{eqnarray*}
F_{K\ell }(\chi)&=&e^{\pm \frac{\alpha_K\chi}{2}} {\tilde \psi}_{K\ell} (\chi),\nonumber\\
{\tilde \psi}_{K\ell}(\chi)&=&  \sin^K\chi
R_{n}^{\alpha_K,\beta_K}(\cot\chi),\nonumber\\
\quad K=n+\ell, &\quad& \alpha_K=\frac{2b}{K+1}, \quad \beta_K=-K.
\end{eqnarray*}
Here, $R_{n}^{\alpha_K,\beta_K}(\cot\chi)$ are the Romanovski polynomials,
which are obtained by means of the Rodrigues formula from the  weight function,
\begin{equation}
  \omega^{\alpha,\beta}(x)=(1+x^2)^{\beta-1} \exp(-\alpha \cot^{-1}x),
   \end{equation}
(see  \cite{Raposo} for a review).

As already mentioned at the end of section \ref{sec:chargeneu}, the cotangent function defining the potential in (\ref{Frnds3}) relates to the fundamental solution to the Laplace-Beltrami operator on $S^3$.
These functions are well known to share the symmetry of the Laplace-Beltrami operator, which in the case under consideration is the four-dimensional rotational group, $SO(4)$, the isometry group of the  $S^3$ manifold. However,
the wave functions in (\ref{sol_CCl}) happen to depend, through the parameters of the Romanovski polynomials, on the representation index $K$, the value of the four-dimensional angular momentum. For this reason, as we shall see below in appendix \ref{sec:appendix2},  
the anticipated $SO(4)$ symmetry of the interaction will be representation dependent, 
much like the dynamical $SO(4)$ symmetry of the Hydrogen atom.

Notice that the sign in front of the cotangent term is not essential, because it depends on the 
orientation chosen on the sphere, and can be switched to its opposite by changing  the range of 
values taken by the  $\chi$ variable  from $\chi\in [0,\pi]$, to  $\chi\in [\pi, 0]$. The point 
is that the function, $r(\chi)=e^{c\chi}$, is the logarithmic spiral and the sign of the 
constant $c$ only defines whether  the points on the curve rotate along a left  or 
right-handed  spiral.

Let us remark that the quasi-radial parts  ${\widetilde \psi}_{K\ell}(\chi)$ of the $\Phi_{K\ell ,m}(\chi, \theta, \varphi)$ functions above allow for the following finite decomposition in the basis of the $S_{K\ell}(\chi)$ part of the ultra-spherical harmonics \cite{RiveraMK}, 

\begin{eqnarray}
   {\widetilde \psi}_{K\ell }(\chi)&=&
   \sum_{\ell^\prime =\ell}^{K} C_{K\ell}^{\ell^\prime}S_{K\ell^\prime}(\chi).
\label{dem3}
\end{eqnarray}

The coefficients $C_{K\ell}^{\ell^\prime}$  specific to  the decomposition in
(\ref{dem3}) have been calculated for several cases and reported in \cite{RiveraMK}. 
In matrix form, the latter equation reads,

\begin{eqnarray}
  {\widetilde {\boldsymbol{\psi}} }_K(\chi)&=&{\mathbf A}^{(K)}{\mathbf S}_K(\chi),\nonumber\\
    {\widetilde {\boldsymbol{\psi}}}_K(\chi)=
    \left(
    \begin{array}{c}
{\widetilde \psi}_{KK}(\chi)\\
{\widetilde \psi}_{K(K-1)}(\chi)\\
.\\
{\widetilde \psi}_{K\ell}(\chi)\\
.\\
{\widetilde \psi}_{K0}(\chi)\\
  \end{array}
  \right), &\quad&
{\mathbf  S}_K(\chi)=
\left(\begin{array}{c}
{ S}_{KK}(\chi)\\
{S}_{K(K-1)}(\chi)\\
.\\
{S}_{K\ell}(\chi)\\
.\\
{ S}_{K0}(\chi)\\
  \end{array}
  \right), 
\label{basis_change}
\end{eqnarray}
where ${\mathbf A}^{(K)}$ is the $(K+1)\times (K+1)$ matrix whose elements are the constants $C_{K\ell}^{\ell'}$. i.e.
\begin{equation}
{\mathbf A}^{(K)}_{\ell \ell^\prime}=C_{K\ell}^{\ell'}.
\label{A_ma_el}
\end{equation}

Returning now to the complete four-variable equation in  (\ref{we3}) with
${\mathcal K}(\chi,\theta,\varphi)$ from (\ref{C3}), 
and with  (\ref{dAl1_1}) in mind, the decompositions in (\ref{dem3})  allow to
conclude on the equivalence between (\ref{Frnds3}) and 
\begin{eqnarray}
  &&\left[
    \frac{\partial^2}{\partial\tau^2}-\frac{\partial^2}{\partial\chi^2} - 2\cot\chi\frac{\partial}{\partial \chi} +\frac{{\bf L}^2(\theta,\varphi)}{\sin^2\chi}
    +1\pm \alpha_K(K+1)\cot\chi +\frac{\alpha_K^2}{4}   \right]\nonumber\\
  &\times&\left[e^{i(K+1)\tau } e^{\pm \frac{\alpha_K\chi}{2}}\sum_{\ell^\prime =\ell}^{K} C_{K\ell}^{\ell^\prime}S_{K\ell^\prime}(\chi)Y_{\ell^\prime}^m(\theta,\varphi)\right] =0.
\label{dem33}
 \end{eqnarray}
Stated differently, the zero-eigenvalue of the operator on the left hand side in   ~(\ref{Frnds3}) is conserved by both bases in (\ref{sol_CCl}) and (\ref{dem33}),  
so implying (in accord with \eqref{dAl1_1}) the interchangeability among,

\begin{equation}
 {\widetilde \psi}_{K\ell }(\chi)Y_\ell^m(\theta,\varphi)
\leftrightarrow
\sum_{\ell^\prime =\ell}^{K}   C_{K\ell}^{\ell^\prime}
S_{K\ell^\prime }(\chi)Y_{\ell^\prime}^m(\theta,\varphi).
  \label{bss_eqv}
\end{equation}
a reason for which the two sets of functions  can be considered equivalent.
This observation will acquire importance in what follows.

 Finally, upon changing, analogously to (\ref{var_chng}), the quasi-radial function,
${\tilde \psi}_{K\ell}(\chi)$,  as
\begin{equation}
{\tilde \psi}_{K\ell}(\chi)=\frac{U_{K\ell}(\chi)}{\sin\chi}
\label{o1}
\end{equation}
the following reduced one-variable  Schr\"odinger equation,
\begin{eqnarray}
\left[
    -\frac{
      {\mathrm d}^2}
    {{\mathrm d}\chi^2} +\frac{\ell(\ell+1)}{\sin^2\chi} \pm \alpha_K(K+1)\cot\chi+\frac{\alpha_K^2}{
      {4}}+1 \right]
e^{\pm \frac{\alpha_K\chi}{2}} U_{K\ell}(\chi)&&\nonumber\\
=\left[(K+1)^2 +1\right] 
         e^{\pm \frac{\alpha_K\chi}{2}}U_{K\ell}(\chi),&&
\label{cot_1DSchr}
\end{eqnarray}
emerges as an extension of \eqref{SchrScrf} by a cotangent term.

\section{Turning on the conformal potential on $S^1\times S^3$ through conformal  metric deformations}\label{sec:harmpot}

Here we show that through the conformal deformation,
\begin{equation}
  g^\ast =e^{-2f}g, \quad f=\frac{\alpha_K\chi}{2},
  \label{dem_pre}
\end{equation}
of the round  metric,
$g$  in (\ref{Ads52}) corresponding to ${\mathcal M}^{1,3}\simeq S^1\times S^3$,
with  $\alpha_K(K+1)=2b$, and $b$  in (\ref{Frnds3}),
the $(2b\cot \chi +\alpha_K^2/4) $ potential in (\ref{Frnds3})
can be introduced along the lines of (\ref{Canz23}) into
the $\Box_{S^1\times S^3} (\tau,\chi)$ operator given in  (\ref{instr1}).

\subsection{The particular case of the ground state}
Indeed, subjecting (\ref{instr1}) to the  conformal  re-scaling in (\ref{dem_pre}) for $K=0$
amounts to,

\begin{eqnarray}
  \Box_{(g^\ast=e^{-\alpha_0\chi}g)}(\tau,\chi)&=&e^{2(\frac{\alpha_0\chi}{2}-\tau)}
  e^{\frac{\alpha_0\chi}{2}}\Box_{S^1\times S^3}(\tau,\chi) e^{\frac{-\alpha_0\chi}{2}}\nonumber\\
  &=&e^{2(\frac{\alpha_0\chi}{2}-\tau)}\left[
    \Box_{S^1\times S^3}(\tau,\chi) + D_{0} -\frac{\alpha_0^2}{4} +
  \alpha_0\cot\chi\right]\,,
\label{t_gst}
\end{eqnarray}
where 
\[
D_0=\alpha_0 \frac{\partial }{\partial \chi}\,.
\]
In this way, the re-scaled $\Box_{(g^\ast=e^{-\alpha_K\chi }g)}(\tau,\chi)$
operator takes the form of an extension of $\Box_{S^1\times S^3}(\tau,\chi)$ in (\ref{instr1})  by a  potential, call it $V_0(\chi)$, 
\begin{equation}
V_0(\chi)=\alpha_0\cot\chi -\frac{\alpha^2_0}{4} +\alpha_0 \frac{\partial}{\partial \chi},
\label{pt_Pt}
\end{equation}
picking up the positive sign for the magnitude of the cotangent function.

Making now use of (\ref{fund_thrm}) (for $n=4$), and taking into account (\ref{qsrad}),
we conclude that,

\begin{eqnarray}
\left[\frac{\partial ^2}{\partial \tau ^2} 
  -\frac{\partial^2}{\partial \chi^2} -2\cot\chi \frac{\partial }{\partial \chi}
  + D_{0} -\frac{\alpha_0^2}{4} +
  \alpha_0\cot\chi  +1 \right]\nonumber
   \left[e^{i\tau}e^{\frac{\alpha_0\chi}{2}}S_{00}(\chi)\right]=0\,. 
  \end{eqnarray}
This equation is indeed satisfied, because substitution of
\begin{equation}
  \alpha_0\frac{\partial }{\partial \chi }e^{\frac{\alpha_0\chi}{2}}=\frac{\alpha_0^2}{2}e^{\frac{\alpha_0\chi}{2}}, \end{equation}
into (\ref{pt_Pt}) gives,

\begin{eqnarray}
  &&\left[ \Box_g(\tau,\chi) +\alpha_0\cot \chi +
    \frac{\alpha_0^2}{4}\right]
  \left[e^{i\tau}e^{\frac{\alpha_0\chi}{2}}S_{00}(\chi)\right]\nonumber\\
&=&e^{2(\frac{\alpha_0\chi}{2}-\tau)}\left[\frac{\partial ^2}{\partial \tau ^2} 
  -\frac{\partial^2}{\partial \chi^2} -2\cot\chi \frac{\partial }{\partial \chi }
  + \alpha_0\cot\chi  + \frac{\alpha_0^2}{4} +1\right] \left[e^{i\tau}e^{\frac{\alpha_0\chi}{2}}S_{00}(\chi)\right]\\
  &=&e^{2(\frac{\alpha_0\chi}{2}-\tau)}\left[ -1^2 + 1^2\right] \left[e^{i\tau}e^{\frac{\alpha_0\chi}{2}}
    S_{00}(\chi)\right]=0\,, \nonumber
\label{HMWRK_1}
\end{eqnarray}
where we made use of (\ref{Frnds8}).  Upon  the identification,  $\alpha_0=2b$,
we observe that as expected,  the  equation (\ref{HMWRK_1}) has the same form as (\ref{Frnds3}), picking up the $2b\cot\chi$ term with a positive sign.  Therefore, the corresponding ground state solution in (\ref{sol_CCl}) could be recovered.
 
\subsection{The case of an arbitrary $K$}\label{subsec:arbitraryk}

We repeat the procedure from the previous section for the case of an arbitrary $K$, 
by writing,

\begin{eqnarray}
  { \Box}_{(g^\ast =e^{-\alpha_K\chi}g)}(\tau, \chi) &=&
  e^{2\left(\frac{\alpha_K\chi}{2} -\tau \right)} \, e^{\frac{\alpha_K\chi}{2}}\Box_g(\tau, \chi) e^{-\frac{\alpha_K\chi}{2}} \,.
\label{dem1}  \nonumber\\
\end{eqnarray}
Dragging the $e^{-\frac{\alpha_K\chi}{2}}$ factor from the very right to the very left, amounts to,

\begin{equation}
 e^{2(\frac{\alpha_K\chi}{2}-\tau )} \, e^{\frac{\alpha_K\chi}{2}}\Box_g(\tau, \chi) e^{-\frac{\alpha_K\chi}{2}} =
e^{2(\frac{\alpha_K\chi}{2}-\tau)}\left[\Box_{S^1\times S^3}(\tau,\chi) + D_{K} -\frac{\alpha_K^2}{4} +\alpha_K(K+1)\cot \chi\right]\,,
  \label{dem2}
\end{equation} 
where
\[
D_K =\alpha_K\left(\frac{\partial }{\partial \chi}-K\cot\chi\right)\,.
\]

Similarly to the special case of $K=0$ considered above, in general, the conformal wave operator associated to the deformed metric takes the form of an extension of $\Box_{S^1\times S^3}(\tau,\chi)$  by  a potential, call it ${ V}_K(\chi)$, which generalizes  (\ref{pt_Pt})  and is  given by,
\begin{equation}
{ V}_K(\chi)=\alpha_K(K+1)\cot \chi + \alpha_K\left(
  \frac{\partial }{\partial \chi}-K\cot\chi\right) -\frac{\alpha_K^2}{4} .
\label{gen_ind_pt}
\end{equation}
Now we are searching for $e^{i(K+1)\tau }u_{K\ell}(\chi)$ functions which make (\ref{dem1}) vanish. Assuming the basis in the right-hand-side in (\ref{bss_eqv}) as the solution of (\ref{dAl1}),
and making use of \eqref{fund_thrm}, we need to prove that $e^{i(K+1)\tau} u_{K\ell}(\chi)$, with $u_{K\ell}(\chi)$ taken as
\begin{equation}
  u_{K\ell}(\chi)=e^{\frac{\alpha_K\chi}{2}}\sum_{\ell^\prime =\ell}^{K} C_{K\ell}^{\ell^\prime} S_{K\ell^\prime }(\chi),
\label{new}
\end{equation}
renders  (\ref{dem2}) equal to zero.

Indeed, the substitution of (\ref{new}) into (\ref{dem2}, and upon accounting for  (  \ref{point_1}), leads to,

\begin{eqnarray}
&&  {\Box}_{(g^\ast =e^{-\alpha_K\chi}g)}(\tau, \chi)\left[  e^{i(K+1)\tau}e^{\frac{\alpha_ K\chi}{2}}
  \sum_{\ell^\prime =\ell}^{K} C_{K\ell}^{\ell^\prime} S_{K\ell^\prime }(\chi)\right]\nonumber\\
&=&e^{2\left(\frac{\alpha_K\chi}{2} -\tau\right)}\left[\frac{\partial ^2}{\partial \tau ^2} 
  +{\mathcal K}(\chi) +1 + D_{K} -\frac{\alpha_K^2}{4} +\alpha_K(K+1)\cot \chi \right]\nonumber\\
  &\times& \left[
    e^{i(K+1)\tau} e^{\frac{\alpha_K\chi}{2}}\sum_{\ell^\prime =\ell}^{K} C_{K\ell}^{\ell^\prime}
    S_{K\ell^\prime }(\chi)\right]\nonumber\\
  &=&e^{2\left(\frac{\alpha_K\chi}{2}-\tau\right)}e^{i(K+1)\tau}\left[-(K+1)^2  -\frac{\partial^2}{\partial \chi^2} -2\cot\chi \frac{\partial }{\partial \chi} +\alpha_K(K+1)\cot\chi +1\right]\nonumber\\
  &\times&\left[e^{\frac{\alpha_K\chi}{2}}\sum_{\ell^\prime =
\ell}^{K} C_{K\ell}^{\ell^\prime} S_{K\ell^\prime m}(\chi)\right]
  +e^{2\left(\frac{\alpha_K\chi}{2} -\tau\right)}e^{i(K+1)\tau}e^{\frac{\alpha_K\chi}{2}}\nonumber\\
  &\times&\left[
\frac{\ell (\ell +1)}{\sin^2\chi} +D_K -\frac{\alpha_K^2}{4}+\frac{\alpha_K^2}{2}
\right]\left[ \sum_{\ell^\prime =\ell}^{K}C_{K\ell}^{\ell^\prime} S_{K\ell^\prime}(\chi)\right]\label{dem4}\\
  &=&e^{2\left(\frac{\alpha_K\chi}{2}-\tau\right)}e^{i(K+1)\tau}  \sum_{\ell^\prime =\ell}^{K}\left[-(K+1)^2
     +{\mathcal K}(\chi)+1+\alpha_K(K+1)\cot \chi
+\frac{\alpha_K^2}{4} \right]\nonumber\\
  &\times&  \left[e^{\frac{\alpha_K\chi}{2}} C_{K\ell}^{\ell^\prime} S_{K\ell^\prime }(\chi)\right].\label{Al1}
  \end{eqnarray}
The justification for the step from (\ref{dem4}) to (\ref{Al1}) 
can be found in  \cite{RiveraMK}. Making now use  of (\ref{we3}) in combination with (\ref{Frnds3}) and (\ref{Frnds8}), amounts to,

\begin{eqnarray}
&&\sum_{\ell^\prime =\ell}^{K}\left[  -\frac{\partial^2}{\partial \chi^2} -2\cot\chi \frac{\partial }{\partial \chi}
    +\frac{\ell^\prime (\ell^\prime +1)}{\sin^2\chi}+1+\alpha_K(K+1)\cot \chi
    \right]\nonumber\\
  &\times&\left[e^{\frac{\alpha_K\chi}{2}} C_{K\ell}^{\ell^\prime} S_{K\ell^\prime }(\chi)
    \right]\nonumber\\
   &=&\left( (K+1)^2 -\frac{\alpha_K^2}{4}\right)\left[e^{\frac{\alpha_K\chi}{2}}\sum_{\ell^\prime =\ell}^{K} C_{K\ell}^{\ell^\prime} S_{K\ell^\prime }(\chi)\right],\label{Schr_my}
  \end{eqnarray}
which, substituting in (\ref{Al1}), leads to

\begin{eqnarray}
  &&  {\Box}_{(g^\ast=e^{-\alpha_K\chi }g)}(\tau, \chi)\left[ e^{i(K+1)\tau}e^{\frac{\alpha_ K\chi}{2}}\sum_{\ell^\prime =\ell}^{K} C_{K\ell}^{\ell^\prime} S_{K\ell^\prime }
    (\chi)\right]\nonumber\\
  &=&e^{2(\left(\frac{\alpha_K\chi}{2}-\tau\right)}\left[
    -(K+1)^2 +(K+1)^2 -\frac{\alpha_K^2}{4} +\frac{\alpha_K^2}{4} \right]\nonumber\\
  &\times& \left[ e^{i(K+1)\tau}e^{\frac{\alpha_K\chi}{2}}\sum_{\ell^\prime =\ell}^{K} C_{K\ell}^{\ell^\prime} S_{K\ell^\prime }(\chi)\right]=0. \label{Al3}
\end{eqnarray}

Comparison of the last two terms in the squared brackets in  (\ref{Al1}), to the {\it ad hoc} potential  in (\ref{Frnds3}), reveals the relation between the former's and the fundamental solution of the Laplacian on $S^3$.
In fact, through the deformation of the $S^1\times S^3$ metric in (\ref{dem_pre}), the potential by which we extended (\ref{instr1})
to (\ref{Frnds3}), hereafter denoted by
\begin{equation}
  V_{CCD}(\chi)=\alpha_K(K+1)\cot\chi +\frac{\alpha_K^2}{4},
\label{Hrm_PT}
\end{equation}
(with $\alpha_K$ as in (\ref{sol_CCl})), could be generated through a conformal deformation of the  $S^1\times S^3$ metric.
Therefore, the source of the potential lies in the new curvature.

Thus, the conclusion can be made that at the le\-vel of the wave equation (\ref{Al3}), at which $e^{2\left(\frac{\alpha_K\chi}{2} -\tau\right)}$ can be dropped, free motion 
 on the con\-for\-mal\-ly re\-scal\-ed ${\mathcal M}^{1,3}\simeq S^1\times S^3$ metric 
 $g^\ast$, is equivalent to motion within the $V_{CCD}(\chi)$  potential on the ini\-ti\-al round met\-ric $g$, a potential which, as we have seen, corresponds to a charge dipole.
For this reason, we have called it `con\-fo\-rmal charge dipole' \-$(CCD)$ po\-ten\-ti\-al.

Conformal transformations on $S^p\times S^q$ give rise \cite{Scho} to the  maximal conformal algebras  $so(p+1,q+1)$ of dimensionality $(p+q+1)(p+q+2)/2$, being
$p=1$, $q=3$, and $so(2,4)$  the case of our interest. Indeed, the  solutions to (\ref{Frnds3})  in equation (\ref{bss_eqv})
are typical functions contained in the representation induced by the metric deformation in (\ref{dem_pre}).
This representation is constructed in Appendix \ref{sec:appendix2} as an intertwining transformation of the generators of the isometry group of $S^1\times S^3$ listed in the Appendix \ref{sec:appendix1}.

\section{Phenomenology of color confinement from the compactified Min\-kow\-ski spacetime}\label{sec:phenomenology}

We will argue that for a proper choice for $\alpha_K$ , anyone of the two equations,
\begin{eqnarray}
  \Box_{(g^\ast =e^{-2f}g)}(\tau,\chi,\theta,\varphi)\psi (\tau,\chi,\theta,\varphi)&=&0,
\label{pstj_1}
\end{eqnarray}
 or
\begin{eqnarray}
  \left(-\Delta_{S^3}(\chi,\theta,\varphi) +V_{CCD}(\chi) \right)\Phi(\chi,\theta,\varphi)
 & =&(K+1)^2\Phi(\chi,\theta,\varphi),
\label{Pstj_2}
\end{eqnarray}
is suited as  a realistic quantum mechanical  limit to QCD in the infrared.

The conformal charge dipole  potential in (\ref{Hrm_PT}), as 
incorporated \emph{ad hoc} on ${\mathcal M}^{1,3}\simeq S^1\times S^3$ by means of
\eqref{Frnds3}--\eqref{sol_CCl}, and into the regular Schr\"odinger framework  by means of 
\eqref{cot_1DSchr}, has already found systematic applications
in the description of those properties of hadrons,  both mesons and baryons, which reflect, by  opening the conformal window in the infrared \cite{Deur}, the underlying  conformal symmetry of the strong interaction in this regime, although its deep connection to the compactified Minkowski spacetime and the ${\mathbf R}^{2,4}$ null-ray cone, i.e. the conformally compactified $\mathrm{AdS}_5$ boundary, seems to have been overlooked. 

The hadron mass distributions with total angular momentum and parity (hadron spectra) have to be mentioned. These distributions are clearly dominated by the quantum numbers of the irreducible finite dimensional representations of the  $SO(4)$ group, an observation frequently emphasized by various authors, \cite{KC2016} and references therein. In particular, in that  work, the spectra of mesons belonging to the  four families,  $\pi$, $f_0$, $a_0$, and $\eta$ (a total of 71 mesons)  have been investigated in detail, and fairly well explained by means of a potential based on (\ref{dem33}), fixing the compactification radius to $R=0.58$ fm. Also in that work, arguments are given in favor of expressing the potential magnitude, denoted by $2b$ through the present text, as
\begin{equation}
2b=\alpha_sN_c,
\label{alphas_Nc}
\end{equation}
with $\alpha_s$ and $N_c$ standing for the strong coupling constant, and the number of colors,
respectively. Within this new physical context, the charge has been identified with the color-electric charge. In fact, the potential relevant to strong interaction is modeled after,
\begin{equation}
  V_{CCD}(\chi)=-\alpha_sN_c \cot\chi,
  \label{CCCD_PT}
    \end{equation}
and, {\it in this parametrization\/}, has been called a `color confining dipole' potential. In its series expansion \cite{KC2016},
\begin{equation}
-\alpha_sN_c \cot\chi =-\frac{\alpha_c N_c}{\chi } +2\alpha_sN_c \chi +..., \quad \chi=\frac{\stackrel{\frown} {r}}{R},
\label{Crnl_agn}
\end{equation}
one finds as a first term the Coulomb-like inverse `distance' (in the arc
$\stackrel{\frown}{r}$ from the North pole)
potential, brought about by one gluon exchange, followed by a linear term in the arc, the expression of strings of gluons causing confinement. This feature establishes a close link between $V_{CCD}(\chi)$ and the phenomenological Cornell potential of a wide spread. 
More specifically, the parametrization in (\ref{CCCD_PT})  appears when
calculating  the potential from two cusped Wilson loops \cite{Belitsky},
$\Gamma_N(\chi)$, and $\Gamma_S(\chi)$, with the opposite  charges being  placed on two antipodal points, $N$ and $S$, of $S^3$, and related to the fundamental solution of the Laplace-Beltrami operator on $S^3$ \eqref{cotsolfund} as,
\begin{eqnarray}
  \Gamma_N(\chi)&=&\frac{\alpha_s N_c}{\pi} (\pi -\chi)\cot\chi,
  \label{cspd_N}\\
  \Gamma_S(\chi)&=&-\frac{\alpha_s N_c}{\pi}\chi\cot\chi.
  \label{cspd_S}
\end{eqnarray}
The emerging dipole potential,
\begin{equation}
\Gamma_S(\chi)-\Gamma_N(\chi)=-\alpha_cN_c\cot\chi,
\label{Al_Bel}
\end{equation}
is then precisely the one in (\ref{CCCD_PT}). However, in
\cite{KC2016} the relevance of $S^3$  for the equation  (\ref{Frnds8}), equivalently (\ref{Pstj_2}), interpreted as the non-relativistic stationary Schr\"odiner equation, 
has been motivated by the assumed validity of $dS_4$ special relativity
whose space-like region contains a hyper-spherical surface of simultaneity, 
thought of as the internal space of the color-charge degrees of freedom.
 Instead, in the present work, the internal spacetime has been identified with the 
 compactified null-ray cone of $\mathrm{AdS}_5$, a procedure which invokes  
 the conformal time via (\ref{Canz1}) thus giving (\ref{Pstj_2}) 
 the interpretation of a Klein-Gordon equation.
For practitioners both options  should compare  in quality. Indeed, as already discussed in section 5, equations (\ref{Ads53}) and(\ref{C1}), the $dS_4$, $AdS_4$, and the Minkowski spacetimes are all related to each other by metric deformations
\citep{Huguet,Canzani}, a reason for which they all feature the traits of conformal symmetry, and present themselves as a  Special Relativity Triple \cite{Zhou}. The approach promoted in the present work  has the advantage over $dS_4$  of  possessing the full set of $so(2,4)$ generators, while on $dS_4$ some of them are missing.

 Along these lines, the number of parameters in the quantum mechanical wave equation employed in meson description  reduces to essentially one, which is the hyper-spherical (compactification)  radius, $R$. In \cite{Ahmed} the spectra of the heavy flavored mesons have been evaluated within the same scheme, achieving fine agreement with data. The compactification radius for the charmonium has been found as $R_{(c\bar c)}=0.56$ fm, while for the bottomonium it is significantly lower, $R_{(b\bar b)}=0.29$ fm. Comparison to the value of $R=0.58$ fm,  relevant for the unflavored mesons \cite{KC2016}, indicates that within the wide range between the pion and charmonium masses, $R$ remains  practically same. 
In \cite{MK_Formfactors} usefulness of the potential under discussion,
included into a Dirac equation on $S^3$, has been demonstrated by
the very satisfactory evaluation of the electromagnetic properties of the nucleon, in first place the proton's and neutron's electric-charge-- and the magnetic-dipole form factors, including their ratio, and together  with the related  mean square radii. Quite recently, the thermodynamic properties of same potential have been worked out in \cite{AramDavid} for the case of quantum meson gases. There, it has been demonstrated that the critical temperature at which a very hot charmonium Bose gas, as it can be  produced at very high temperatures in heavy ion collisions, undergoes a  phase transition to a Bose condensate,  hits Hagedorn's temperature for the $R$ value previously adjusted to the spectrum in \cite{Ahmed}.  
The above results, obtained  by means of the color confining dipole potential in (\ref{CCCD_PT}), seem to be quite encouraging, and speak in favor of the relevance of the boundary of the  compactified $AdS_5$ spacetime to the physics of hadrons and the color confinement.

Moreover, the radius $R$ of the  compactified  Minkowski space time, if used along 
$\Lambda_{QCD}$ as an additional scale in QCD, could be used to reparametrize the transferred momentum $Q^2$ according to, 
\begin{eqnarray}
  Q^2c^2+\frac{\hbar^2 c^2}{R^2 }> \Lambda_{QCD}^2 \mbox{ and hence } R^2<\frac{\hbar^2 c^2}{\Lambda_{QCD} ^2 }\,.
\label{PT_IR}
\end{eqnarray}
Then, the formula for the strong coupling constant $\alpha_s(Q^2)$
corresponding, say, to 1--loop approximation for simplicity, 
\begin{equation}
  \frac{\alpha_s(Q^2)}{\pi}= \frac{4}{\beta_0 \ln \left(\frac{
      Q^2c^2}{\Lambda^2_{QCD}} \right)}, \quad \beta_0=11-\frac{2}{3}n_f,
\label{PI_IR1}
\end{equation}
where  $n_f$ is the number of flavors, would be rewritten as,
\begin{equation}
  \frac{\alpha_s(Q^2)}{4\pi}\approx \frac{1}{\beta_0 \ln \left(
    \frac{
      Q^2 c^2}{\Lambda^2_{QCD}} +\frac{\hbar^2c^2}{R^2\Lambda_{QCD}^2} \right)},
\label{PI_IR2}
\end{equation}
and stops being divergent at $Q^2c^2=\Lambda_{QCD}^2$. In consequence, the strong coupling begins walking (`freezing out') toward the fixed value of,
\begin{equation}
  \frac{\alpha_s(0)}{\pi}= \lim_{Q^2\to 0} \frac{4}{\beta_0 \ln \left(\frac{
      Q^2 c^2}{\Lambda^2_{QCD}} +
    \frac{\hbar^2c^2}{R^2\Lambda_{QCD}^2} \right)}
  \to \frac{4}{\beta_0 \ln \left(\frac{\hbar^2c^2}{R^2\Lambda_{QCD}^2} \right)}.
\label{PI_IR3}
\end{equation}

\begin{figure}[h]
  \begin{center}
  \includegraphics[width=9.5cm]{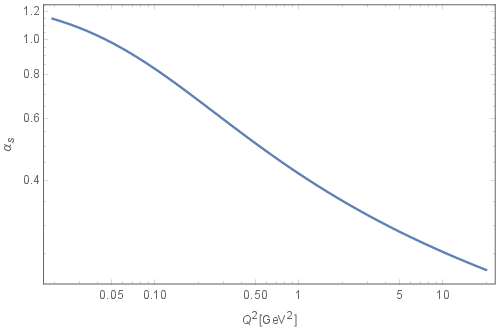}
  \caption{Walking of the strong coupling to a fixed value at origin according to (\ref{PI_IR3}).}
\end{center}
  \end{figure}

A similar scheme for avoiding the logarithmic divergence of $\alpha_s(Q^2)$ has been considered in \cite{Shirkov}, where $Q^2$ has been extended by some squared mass, $m_g^2$, attributed to gluons. In our approach, we relate the reparametrization of $Q^2$ to the more fundamental scale of the finite radius of the compactified $\mathrm{AdS}_5$ boundary, whose value can be fitted to data. In particular,
with the overall $R=0.58$ fm, extracted from fitting light meson spectra  in \cite{KC2016}, 
$\Lambda_{QCD}\approx 200 $ MeV, and for $n_f=3$,  one estimates $\alpha_s(0)$ in Fig.~1 as,
\begin{equation}
\frac{\alpha_s(0)}{\pi }\to 0.42,
\end{equation}
a number which is of same order of magnitude as  the experimentally observed one,
that is, $\lim _{Q^2\to 0}\frac{\alpha_s(Q^2)}{\pi }\to 1$, and reported in \cite{Deur}.

As already mentioned above, the $R$ value for the heavy-flavored mesons reported in \cite{Ahmed} is $R=0.56$ fm, practically the same as the one for light mesons, thus hinting on a certain universality of the  $\mathrm{AdS}_5$ boundary radius of compactification, within the wide range of transferred momenta varying from zero to $\sim 3000$ MeV/$c$. Our estimate is in line with the outcomes of more technical approaches to the `freezing' of $\alpha_s$, reviewed in \cite{Guy}.

\section{Conclusions}\label{sec:conclusions}
In the present work we explored a possible explanation of the non-observability of free color-electric charges  by an internal space-time of $S^1\times S^3$ geometry, which we interpreted as the conformally compactified null-ray cone of ${\mathbf R}^{2,4}$, at the  boundary of $\mathrm{AdS}_5$, also known as the compactified Minkowski spacetime, ${\mathcal M}^{1,3}$. As long as on closed manifolds, color-charge dynamics
can be formulated only for neutral $2^n-$poles, we concluded that this geometry represents a stage suitable for the description of the color-charge neutral mesons as color-electric charge  dipoles, i.e. as quark--anti-quark systems, as they are. The functional shape of the potential generated by such a charge configuration is given  in \eqref{cotsolfund} and
turns out to be related to a fundamental solution of the $S^3$ Laplace-Beltrami operator.
Considering free scalar massless particle on $S^1\times S^3$ in terms of the wave equation (\ref{dAl1}) with the conformal operator in (\ref{we3}),
we then could construct in (\ref{Hrm_PT}) a more precise charge-dipole potential through a conformal deformation of the round  metric, given in (\ref{dem_pre}). Stated differently,  we showed that free motion on the deformed metric in (\ref{dem_pre}) is equivalent to motion on $S^1\times S^3$ in the presence of the potential in (\ref{Hrm_PT}). In subsection \ref{subsec:arbitraryk} we emphasized that the conformal metric deformations induce interactions in a controlled way 
 in the sense that the wave functions solving the potential are determined by the conformal re-scaling factor (\ref{fund_thrm}). This is a serious advantage over wave operators containing  {\it ad hoc} potentials, whose solutions have still to be found, an aspect that may be of practical interest to studies based on numerical simulations.

We furthermore worked out in the Appendix \ref{sec:appendix2} some explicit
expressions\footnote{Admittedly representation dependent, like the one in \eqref{end_answ}.}, given in the form they can appear in dynamical symmetries, for the generators of the conformal algebra on the deformed metric. This is done
with the aid of  the  intertwining  transformation (\ref{smlrt_trnsfrm}) of the  algebra of the $S^1\times S^3$ isometry group, whose generators are listed in Appendix \ref{sec:appendix1}. Such expressions are of interest for spectroscopic studies where they can be used to describe excitation and decay modes of hadrons.

In section \ref{sec:phenomenology} we briefly reviewed earlier applications of this potential to the phenomenology  of hadrons and recalled in (\ref{alphas_Nc})-(\ref{CCCD_PT}), and (\ref{Al_Bel}) its  parametrization in terms of $\alpha_s$ and $N_c$, the fundamental parameters of QCD. This parametrization justifies renaming the interaction in (\ref{Hrm_PT}) from the general `conformal charge dipole', to the more QCD specific `color confining dipole'  potential. 

In fact, in terms of the equation (\ref{Frnds3}) with the parametrization in (\ref{alphas_Nc}) quite a broad range of strong interaction phenomena, ranging from hadron spectra, over electromagnetic form factors, up to phase transitions, could be pretty realistically covered, indeed. For this reason, we consider anyone of the conformal wave equations, the primordial one,
\begin{equation}
  \Box_{\left(g^\ast=e^{-\frac{\alpha_sN_c\chi}{K+1}}g\right)}\psi =0, \quad g:\quad {\mathcal M}^{1,3}\simeq S^1\times S^3,
  \label{Sl_6_8}
  \end{equation}
or its equivalent,
\begin{eqnarray}
  \left[ \Box_{S^1\times S^3} -\alpha_sN_c\cot\chi +
    \frac{\alpha_s^2N_c^2}{4(K+1)^2} \right]\Phi&=&0,
  \label{BiBo}
\end{eqnarray}
as a good  candidate for the description of the quantum mechanical limit of QCD
(\ref{pstj_1}), (\ref{Pstj_2}).
Through the text all the equations have been considered in dimensionless units. In order to convert them in units of MeV$^2$, they have to be multiplied by the left by $\hbar^2c^2/R^2$ (cf. (\ref{PT_IR})).

The equation (\ref{BiBo}) parallels the one used in the evaluation of the Hydrogen atom properties, insofar as it becomes the latter in the $R\to \infty$ limit, and upon replacing the strong interaction parameters $\alpha_s$, and $N_c$ by their  respective electromagnetic $\alpha$, and $ Z$ counterparts. 
One possible explanation for this parallelism is hinted by the observation that  the $so(2,4)$ algebra, generating the isometry group $ SO(2,4)$ at the 
$ AdS_5$ boundary, can be cast in the shape of a graded algebra as
\[
so(2,4)\simeq g_1\oplus [g_1,g_{-1}]\oplus g_{-1}.
\]

Here, $g_1$ and $g_{-1}$ denote two Abelian algebras whose generators are 
the four momentum, $P_\mu$, and the four-vector of special conformal transformations, $K^\mu$, respectively, these being related by an involution,
$  K_\mu=I P^\mu I $, with 
$  I(x_0,-{\mathbf  x })= (x_0/x^2, -{\mathbf x}/x^2)$. Then $[g_1,g_{  -1}]$ recovers the algebra of the Lorentz group with an associated dilaton operator, $  x^\mu \partial_\mu$.

This property of the conformal algebra links it to the Jordan algebra  $J_2^C$ (see \cite{Todor} for details) and vice versa. Along a similar line of reasoning,  the Jordan algebra $J_3^O$ has been employed in \citep{Dubois,Svetla} as a link to the exceptional group $F_4$, hypothesized as a group of unification of strong and electroweak interactions, thus conjecturing its subgroup $G_2$, which can support only colorless systems (now considering three colors) as a possible gauge group of QCD.

Finally, we pointed to the possibility of using, alongside $\Lambda_{QCD}$, the compactification radius $R$ of the $\mathrm{AdS}_5$ boundary as a further scale in QCD, and suggested in (\ref{PI_IR2}) a phenomenological re-parametrization of the one-loop expression for $\alpha_s(Q^2)$, in which the infinity of $\alpha_s\left(Q^2\right)$ at $Q^2c^2=\Lambda_{QCD}^2$ could be removed. Instead, a process of walking of the strong coupling toward a fixed value at $Q^2=0$ starts, thus opening an avenue towards perturbative treatment of processes in the infrared.

In conclusion, we suggest that
it is possible to reach a realistic phenomenological description of the color-confinement in terms of a conformal wave operator equation, whose (very few) potential parameters are directly related to the fundamental constants in QCD.


\appendix
\section{The conformal algebra of the free motion on $S^1\times S^3$}\label{sec:appendix1}
In this appendix we summarize  the properties of  the  
	$so(2,4)$  algebra of the isometry group $SO(2,4)$ of the compactified Minkowski space
	${\mathcal M}^{1,3}\simeq S^1\times S^3$.
	The conformal deformation of the metric 
	$g^\ast =e^{-\alpha_K\chi}g$ in (\ref{dem_pre}) then induces 
	another representation of the very same  $so(2,4)$ algebra.

\subsection{Generators of the conformal algebra on $S^1\times S^3$}

 The generators $L_{ab}$ of pseudo-rotations  of the conformal  algebra 

\begin{equation}
L_{ab}=
-y_a\frac{\partial}{\partial y ^b}+ y_b\frac{\partial}{\partial y^a} 
\ ,
\qquad L_{ab}=-L_{ba},
\label{Jab}
\end{equation}
satisfy the  
$so(2,4)$ commutation relations for the diagonal metric $g_{ab}$ with signature $(++----)$
\begin{equation}
  \left[ L_{ab},L_{cd}\right]=
	g_{ad}L_{bc}+g_{bc}L_{ad} -
    g_{ac}L_{bd}- g_{bd}L_{ac}   \ .
 \label{conf_comm}
\end{equation}

The differential operators $L_{ab}$
can be  expressed in different coordinates of the 6-dimensional projective
space. In particular for the global polar coordinates in $AdS_5$, given by eq.   (\ref{polar_cmpct}), they
can be found in \cite{Bogdanovic}, \cite{Ratiani}, among others.
The seven generators of the {\it maximal } compact subgroup  $SO(2)\times SO(4) \subset SO(2,4)$, namely  $L_{-10}$ for $SO(2)$  and  $L_{ab}$, $1\leq a<b \leq 4$
for SO(4)read

 \begin{eqnarray}
 L_{-10}&=&\partial_\tau,\label{conf_time}\\
 L_{12}&=&\partial_\varphi,\label{so4_gnrts_1}\\
 L_{23}&=&\sin\varphi\partial_\theta +\frac{\cos\theta \cos\varphi}{\sin\theta}\partial_\varphi,\label{so4_gnrts_2}\\
 L_{31}&=&-\cos\varphi\partial_\theta +\frac{\cos\theta\sin\varphi}{\sin\theta}\partial_\varphi,\label{so4_gnrts_3}\\
 L_{14}&=&\sin\theta\cos\varphi \partial_\chi +\cot\chi \cos\theta \cos\varphi 
 \partial_\theta -\cot\chi\frac{\sin\varphi}{\sin\theta} \partial_\varphi,\label{so4_gnrts_4}\\
 L_{24}&=&\sin\theta\sin\varphi \partial_\chi +\cot\chi \cos\theta \sin\varphi
 \partial_\theta +\cot\chi\frac{\cos\varphi}{\sin\theta} \partial_\varphi,\label{so4_gnrts_5}\\
 L_{34}&=&\cos\theta\partial_\chi -\cot\chi \sin\theta\partial_\theta,
\label{so4_gnrts_6}
\end{eqnarray}
while the remaining eight  ones are,
 \begin{eqnarray}
   L_{-11}&=&-\sin\tau \sin\chi\sin\theta  \cos\varphi \partial _\tau +\cos\tau \cos\chi \sin\theta \cos\varphi \partial _\chi\nonumber\\
   &+&\frac{\cos\tau\cos\theta\cos\varphi }{\sin\chi}\partial_\theta -\frac{\cos\tau\sin\varphi }{\sin\chi\sin\theta}\partial_\varphi \\
L_{-12}&=&  -\sin\tau \sin\chi\sin\theta  \sin\varphi \partial _\tau +\cos\tau \cos\chi \sin\theta \sin\varphi \partial _\chi, \nonumber\\
   &+&\frac{\cos\tau\cos\theta\sin \varphi }{\sin\chi}\partial_\theta + \frac{\cos\tau\cos\varphi }{\sin\chi\sin\theta}\partial_\varphi \\
L_{-13}&=& -\sin\tau \sin\chi \cos\theta \partial_\tau +\cos\tau \cos\chi \cos\theta\partial_\chi -\frac{\cos\tau\sin\theta}{\sin\chi}\partial_\theta\\
 L_{01}&=&\cos\tau \sin\chi\sin\theta\cos\varphi \partial_\tau +\sin\tau \cos\chi \sin\theta \cos\varphi\partial_\chi \nonumber\\
&+&\frac{\sin\tau \cos\theta\cos\varphi}{\sin\chi} -\frac{\sin\tau \sin\varphi}{\sin\tau \sin\theta}\partial_\varphi\\
 L_{02}&=&\cos\tau \sin\chi\sin\theta\sin\varphi \partial_\tau +\sin\tau \cos\chi \sin\theta \sin\varphi\partial_\chi \nonumber\\
&+&\frac{\sin\tau \cos\theta\sin\varphi}{\sin\chi} +\frac{\sin\tau \cos\varphi}{\sin\tau \sin\theta}\partial_\varphi,\\
 L_{03}&=& \cos\tau\sin\chi\cos\theta\partial_\tau +\sin\tau\cos\chi\sin\theta \sin\varphi\partial_\chi
- \frac{\sin \tau \sin \theta}{\sin \chi} \partial_\theta\\
L_{-14}&=&-\sin\tau \cos\chi \partial _\tau -\cos\tau \sin\chi \partial\chi\\
 L_{04}&=& \cos\tau\cos\chi\partial_\tau -\sin\tau\sin\chi\partial_\chi.
\label{so24_polar_genrs_9}
 \end{eqnarray}
The dilatation $D$, the Lorentz transformations $L_{\mu \nu}$, the special conformal transformations $K_\mu$, and 
the translations $P_\mu$ are expressed through the conformal generators $L_{ab}$ as follows,
 \begin{eqnarray}
D=-L_{-14}, \quad 
     P_\mu=L_{-1\mu}+L_{\mu 4}, &\quad&
              K_\mu=L_{-1 \mu} -L_{\mu 4} \ ,
\label{DKP}
 \end{eqnarray}
where $\mu, \nu=0,1,2,3$.
 The inverse transformation yields
 \begin{equation}
   L_{-1 \mu} = \frac{1}{2}\left( P_\mu+K_\mu \right), \qquad 
	L_{\mu4} = 	\frac{1}{2}\left(P_\mu-K_\mu \right)
   .
   \label{PpmK}
 \end{equation}
Then the   ``conformal Hamiltonian'' is obtained for $\mu=0$ as
 \begin{equation}
i H:=L_{-10}= \frac{1}{2}\left(P_0+K_0\right)= - i \partial_\tau \ ,
\label{conf_H}
 \end{equation}
and represents a  generator of translation along the compact conformal time $\tau$.

 \subsection{The $so(2,1)$ subalgebra of $so(2,4)$  }

 The compactified time is running  along the circle  $S^1$, the operator of time translation being the conformal Hamiltonian 
\begin{equation}
H=-iL_{-10}.
\label{a3}
\end{equation}
Let us  denote this conformal time  translation by $A_3=H= - i \partial_\tau$. The conformal Hamiltonian $A_3$
 together with the ladder operators
\begin{eqnarray}
 A_{\mp}:=i L_{04}\mp L_{-14} 
 &=&e^{\mp i\tau}\left(\pm \sin \chi \partial _\chi +i\cos\chi \partial _\tau\right),
 \label{Apm}
\end{eqnarray}
close an $so(2,1)$ algebra according to
\begin{eqnarray}
\left[A_3,A_\pm\right]=\pm A_\pm, \qquad \qquad \left[ A_+,A_- \right]&=&-2A_3 \ .
\label{so21}
\end{eqnarray}
The corresponding Casimir $so(2,1)$-operator commutes
with the elements of the  $so(3)$-algebra of the angular momentum operator $L_{ij}$, $1\leq i<j\leq 3$, i.e.
\begin{equation}
A^2=-A_+A_- +A_3^2-A_3 \ .
\label{Casimir}
\end{equation}
The infinite dimensional unitary $so(2,1)$ representations termed to as discrete infinite series, $D^+_j$ and  $D^-_j$ are specified by the basis $\vert j,n\rangle$,
\beq
\ba{ccll}
  A^2\vert j,n\rangle &=&j(j-1)\vert j,n\rangle & \quad j\le 0,\\
  A_3\vert j,n\rangle&=&N\vert j,n\rangle & \quad j\le 0, \quad N=\mp j\pm n, \quad n=0,1,2,\ldots,
	\ea
  \label{dcsr_srs}
\eeq
where the  energies in the $D^+_j$ ($D^-_j$) series are bounded from below (above).
The $so(2,1)$ algebra is the spectrum generating algebra of the free motion on $S^3$ (more details can be found in \cite{Gadella}).
Within the framework of the super-symmetric quantum mechanics  the conformal Hamiltonian $iH\equiv A_3=-i \partial_\tau$
is introduced as an auxiliar operator and its origin from the conformal symmetry remains unrecognized.

\subsection{Generators of $so(4)$ algebra  and 
the spectrum of the free motion  on $S^3$ }

The isometry group of the $S^3$ space is the four-dimensional rotational group, $SO(4)$.
In terms of  shift-operators, the $so(4)$-generators are given by
\begin{eqnarray}
L_{\pm}=L_{23}\pm iL_{31}, &\quad& L_3=L_{12}, \label{so3}\\
N_\pm =L_{14}\pm iL_{24}, &\quad& N_3=L_{34},\label{nvct}
\end{eqnarray}
with $L_{ij}$ for $1\leq i<j \leq 4$ having been defined in (\ref{so4_gnrts_1})-(\ref{so4_gnrts_6}).
The three generators, $L_\pm$, and $L_3$ produce rotations in the coordinate space, while the remaining three, $N_\pm$, and $N_3$ are rotations in the extended euclidean space $\R^4$.

The square of the four-dimensional angular momentum operator
${\mathcal K}$ 
\begin{eqnarray}
  {\mathcal K}&=& {\mathbf L}^2 +
  {\mathbf N}^2,\label{4d_AM}\\
{\mathbf L}^2 &=&\frac{1}{2}\left(L_+L_-+L_-L_+ \right)+L_3^2,\label{so4_s03}\\
{\mathbf N}^2&=&\frac{1}{2}\left(N_+N_-+N_-N_+ \right)+N_3^2,\label{so4_N}
\label{4moment}
\end{eqnarray}
represents one of the  Casimir  $so(4)$-invariants.
The ${\mathbf L}^2$ operator depends on the polar, $\theta$, and azimuthal, $\varphi$, angles in three space, while ${\mathbf N}^2$ depends only on the second polar angle, $\chi$, in the four-dimensional Euclidean space.

The eigenvalue problem of the operator ${\mathcal K}$ in (\ref{4d_AM}) is solved by the ultra-spherical harmonics, 
$Y_{K\ell m}=Y_{K\ell m} (\chi, \theta, \phi)$ as
\begin{equation}
{\mathcal K}Y_{K\ell m}=K(K+2)Y_{K\ell m} \ ,
\end{equation}
where for clarity we  have  suppressed the arguments apparent from the context. 
The eigenvalues of the operator ${\mathcal K}$ in eq. (\ref{4moment})
 are  obtained to be $K(K+2)$,
in agreement with  the spectrum of the free motion on $S^3$ found in  equation (\ref{4Dam}).

\subsection{The $so(2,1)$ 
as spectrum generating algebra of the free motion on $S^3$ }

It is now our goal to review, following \cite{Kusnez},  the property of the  $so(2,1)$ algebra in (\ref{so21}) to serve  as   spectrum generating algebra of (\ref{SchrScrf}). Toward this goal  one first needs to solve the aforementioned equation for a fixed $\ell$ value.
 In changing variables from $\chi$ to $i\cot\chi$, it is not difficult to realize that (\ref{SchrScrf}) transforms (upon a $\sin^2\chi$-factorization) into the differential equation for the associated Legendre function.This observation allows one to write the solution as,
\begin{equation}
\psi_\ell ^N(\chi,\tau )=P_\ell ^N(i\cot\chi)
\label{csc2_sol}
\end{equation}
where $P_\ell^N(i\cot\chi)$ is a generalized  associated Legendre function \cite{Kusnez}. Now one can calculate the action of the operators $A_-$ and $A_+$ on the product of the wave functions in (\ref{csc2_sol}) and the phase factor, $e^{iN\tau}$, i.e. on
 \begin{equation}
  \Psi_\ell ^N(\chi)=e^{iN\tau}P_\ell ^N(i\cot \chi),
  \label{wfau_ttl}
  \end{equation}
 and find,
 \begin{eqnarray}
   A_+ e^{iN\tau} P_\ell^N(i\cot\chi)&=&c^+_{\ell,N}e^{i(N+1)\tau}P_\ell^{N+1}(i\cot\chi),\nonumber\\
   A_- e^{iN\tau} P_\ell^N (i\cot\chi)&=&c^-_{\ell,N}e^{i(N-1)\tau} P_\ell ^{N-1}
   (i\cot\chi),
       \label{shifts}
 \end{eqnarray}
 where $c_{\ell ,N}^\pm$ are some properly defined  constants. In effect, one
 encounters the following eigenvalue problem,

 \begin{equation}
A_+A_- e^{iN\tau}P_\ell^N(i\cot \chi)=\lambda e^{iN\tau}P_\ell ^N(i\cot \chi), \quad \lambda =-(\ell +N)(\ell -N+1).
\label{fctrzt_1}
 \end{equation}

 One now notices that the $A_+A_-$ operator expresses in terms of the $so(2,1)$ Casimir invariant, denoted by $A^2$ as
 \begin{equation}
A_+A_-=-A^2+A_3^2-A_3,
   \end{equation}
 which, upon substitution in (\ref{fctrzt_1}) and accounting for the explicit expression of the Casimir operator in (\ref{Casimir}),
 \begin{equation}
   A^2=\sin^2\chi\left( \frac{\partial ^2}{\partial \chi^2} -\frac{\partial ^2}{\partial \tau^2}\right),
\label{explct_Cas}
 \end{equation}
 amounts to,

 \begin{eqnarray}
\left[
    -\frac{
      \partial^2}
    {\partial \chi^2} +\frac{\ell (\ell +1)}{\sin^2\chi} +1\right]e^{iN\tau}P_\ell^N (i\cot \chi)&=&-\frac{-\partial ^2}{\partial \tau^2}e^{iN\tau }P_\ell^N (i\cot \chi)\nonumber\\
&=&N^2 e^{iN\tau}P_\ell ^N(i\cot\chi),
\label{Gl1}
 \end{eqnarray}
 Upon cancellation of the phase factor, the latter equation is equivalent to (\ref{SchrScrf}). The above considerations reveal the algebra in (\ref{so21})
 as the spectrum generating algebra of the Schr\"odinger equation with the $\csc^2\chi$ potential. In now subjecting the latter equation to the intertwining transformation by the inverse $\sin\chi$ function, one arrives at

 \begin{eqnarray}
   \frac{1}{\sin\chi}
   \left[
    -\frac{\partial^2}
    {\partial\chi^2} +\frac{\ell (\ell +1)}{\sin^2\chi} \right]\sin\chi
   \, \frac{1}{\sin\chi}P_\ell ^N (i\cot\chi)\nonumber&&\\
=\left( {\mathcal K} +1\right) \frac{1}{\sin\chi}P_\ell ^N (i\cot\chi)
=N^2\frac{1}{\sin\chi}P_\ell ^N (i\cot \chi),&&
\label{Gl2}
 \end{eqnarray}
 equation which is equivalent to the description of free motion on $S^3$ according to (\ref{SchrScrf}). Comparison to (\ref{dcsr_srs}) reveals the solutions in (\ref{wfau_ttl}) as $D^+_{\ell +1}$, being the energy values equal to

\begin{equation}
 N^2=(\ell +1 +n)^2=(K+1)^2.
\label{spctr_fr}
\end{equation}

 \noindent
 The equation (\ref{Gl1}) allows also for an alternative  solution expressed in terms of the Gegenbauer polynomials. Indeed, substitution of eqs.~(\ref{var_chng}) and (\ref{qsrad}) into (\ref{Gl1}), followed by a drag of the $\sin^{\ell +1}\chi$ function to the very left, and its subsequent cancellation on both sides, allows to replace (\ref{Gl1}) by the differential equation for the Gegenbauer polynomials  given by,
 \begin{eqnarray}
   \left[
     (1-x^2)\frac{{\mathrm d}^2}{{\mathrm d}x^2} -2(\ell +2) x\frac{{\mathrm d}}{{\mathrm d}x} +n(n+2\ell +2)\right]G_{n}^{\ell +1}(x)=0,&&\nonumber\\
 x=\cos\chi.&&
\label{Gegnb}
 \end{eqnarray}
 Correspondingly, the ladder operators in (\ref{Apm}) in the new basis  become \cite{Levai}
 \begin{equation}
   J_\pm=A_\pm \mp e^{\pm i\tau}(\ell +1)\cos\chi,
   \label{Gbr_basis}
\end{equation}
 and the solutions of (\ref{Gl1}), alternative to those in (\ref{wfau_ttl}), can be written  as,
 \begin{equation}
 {\widetilde \Psi}_{\ell+1}^N(\chi) =   e^{\pm iN \tau}{\sin^{\ell +1}\chi}G_n^{\ell +1}(\cos\chi)= e^{\pm iN \tau}\sin\chi S_{K\ell}(\chi).
\label{alt_bss}
 \end{equation}

 \section{Lie symmetries and algebras for the deformed metric of  $S^1\times S^3$}\label{sec:appendix2}
 This section is devoted to the expressions  of the generators of the
 $so(2,4)$ algebra on the deformed metric in (\ref{dem_pre}), equivalently, 
 of the  equation (\ref{Pstj_2}), in terms of the generators of the
 $so(2,4)$ algebra of the isometry group of $S^1\times S^3$, presented in the preceding appendix.
 Although the conformal symmetry characterizing the round metric
 continues characterizing  also the conformally deformed one, we nonetheless consider it important, for the sake of future applications,  to prove this constructively, and work out  the explicit expressions for the generators in the new representation.

 Recalling that the solutions to the re-scaled conformal wave operator defined in (\ref{dem1}) have been identified as (\ref{new}),
allows to cast (\ref{Al3}) in a matrix form according to,

\begin{eqnarray}
 e^{2(\tau +f)} \,
 e^f\Box_{S^1\times S^3}(\tau,\chi)
      e^{-f}  \left[e^{i(K+1)\tau} e^{f} \sum_{\ell^\prime}{\mathbf A}^{(K)}_{\ell \ell^\prime}
 S_{K\ell^\prime}(\chi)\right]=0,&&\nonumber\\
  \ell,\ell^\prime =0, ..., K,&&
\label{mic_1}
\end{eqnarray}
where  ${\mathbf A}^{(K)}$ are  quadratic $(K+1)\times (K+1)$ matrices in
(\ref{A_ma_el}).  In the notation of (\ref{basis_change}),
\begin{equation}
  e^f{\widetilde {\boldsymbol{\psi}}}_{K}(\chi)=e^f {\mathbf A}^{(K)} {\mathbf S}_{K}(\chi),
\label{bs_trnsfm}
\end{equation}
holds valid.
  For that reason, and ignoring the non-zero factor of $e^{2(\tau +f)}$, the equation in (\ref{mic_1}) simplifies to,
  \begin{equation}
e^f\Box_{S^1\times S^3}(\tau,\chi){\mathbf 1}_{(K+)\times (K+1)}
      e^{-f} \,\,  e^{i(K+1)\tau} e^{f} {\mathbf A}^{(K)}
  {\mathbf S}_K(\chi)=0.
\label{this_eq}
    \end{equation}
  This equation reveals the appearance of the conformal charge dipole  potential $V_{CCD}(\chi)$ in (\ref{Hrm_PT}) as the result of  an intertwining transformation of the  $ {\mathcal K}(\chi,\theta,\varphi)$ part in
$\Box^1_g(\tau,\chi,\theta,\varphi)$ by the $e^f$ function, the square root of the inverse conformal factor defining the deformation of  the metric according to (\ref{mtrc_dfrm}),  as
  \begin{eqnarray}
   e^f{\mathcal K}{\mathbf 1}_{(K+1)\times(K+1)}e^{-f} e^{f} {\mathbf A}^{(K)}
  {\mathbf S}_K
    = \left({\mathcal K}  +\alpha_K(K+1)\cot\chi
    +\frac{\alpha_K^2}{4}\right)
     e^{f} {\mathbf A}^{(K)}
 {\mathbf S}_K\,.
\label{Al2}
  \end{eqnarray}
  As long as ${\mathcal K}(\chi)$ is the (reduced) Casimir operator of the isometry  group $SO(4)$ of the $S^3$ manifold, we are allowed to conclude that the  cotangent interaction on $S^3$ possess same symmetry as the free motion, and that the algebras of the free and interacting particle  cases, when considered in the  $ e^{f} {\mathbf A}^{(K)}{\mathbf S}_K$ basis, are related by an intertwining transformation by the square-root of the inverse conformal factor, the one
  in (\ref{dem_pre}) for our case. In this way, within  the basis ${\mathbf A}^{(K)}{\mathbf S}_K(\chi)$ in (\ref{basis_change}), and the deformation in (\ref{dem_pre}) one finds

  \begin{equation}
    L_{ab}|_{(g^*=e^{-2\frac{\alpha_K\chi}{2}}g)}=e^{\frac{\alpha_K\chi}{2}}L_{ab}|_{g }e^{-\frac{\alpha_K\chi}{2}},\quad a,b=1,2,3,4.
    \label{smlrt_trnsfrm}
  \end{equation}
where $g$ denotes the round $S^1\times S^3$ metric.
  If one opts for choosing as a basis of the free motion on $S^3$ the mere
  ${\mathbf S}_K(\chi)$ columns, then it should be obvious that the ${\mathbf A}^{(K)}$ matrices have to be passed to the intertwining transformation. 

This technique seems to be quite effective. Only in a previous
  work \cite{RiveraMK} by one of us, the $so(4)$ subalgebra of the conformal algebra has been handled  within same scheme. Before, in 1979 Higgs \cite{Higgs79} and Leemon \cite{Leemon79} initiated studies of  the dynamical symmetries of the Kepler problem on the sphere, though they employed the non-conformal gnomonic projection to map the  sphere on a plane, a projection which is neither angle-, nor area preserving. With such a strategy, the conformal algebra  remained obviously beyond reach, and only  a deformed part of it could be detected.
 
In the next subsection we illustrate for practical purposes  the construction of one of the $so(2,1)$ generators following these ideas.

\subsection{The spectrum generating algebra $so(2,1)$}

As an illustrative example for an $so(2,1)$ generator, we
here construct the ladder operator in the transition,
\begin{equation}
F_{11}(\chi) \to
F_{21}(\chi),
\label{so12_1}
\end{equation}
where
\begin{eqnarray}
  F_{11}(\chi)&=&e^{\frac{\alpha_1\chi}{2}}{\widetilde \psi}_{11}(\chi),\nonumber\\
  F_{21}(\chi)&=&e^{\frac{\alpha_2\chi}{2}}{\widetilde \psi}_{21}(\chi),
\end{eqnarray}
are solutions  (\ref{sol_CCl}) to the interacting particle  case in (\ref{Frnds3})-(\ref{Frnds8}). They belong  to levels whose principal quantum numbers are distinct by one unit, i.e. $\Delta K=1$, while the orbital angular momentum has been kept the same, $\ell =1$, in the case under consideration. These are precisely the selection rules imposed by the $so(2,1)$ ladder operators. Thus, the key point of this task concerns the  construction of the 
$\widetilde{\psi}_{11}(\chi)\to {\widetilde \psi}_{21}(\chi)$ ladder operator. To find it, we proceed in the following way: According to \cite{RiveraMK} the following decompositions in the
(\ref{basis_change}) basis  hold valid,

\begin{eqnarray}
{\widetilde \psi}_{11}(\chi)=S_{11}(\chi),&&
{\widetilde \psi}_{10}(\chi)=S_{10}(\chi)+bS_{11}(\chi),\nonumber\\
{\widetilde \psi}_{22}(\chi)&=&S_{22}(\chi)=\sin\chi S_{11},\nonumber\\
  {\widetilde \psi}_{21}(\chi)&=&S_{11}(\chi) +\frac{2b}{3}
    S_{12}(\chi),\nonumber\\
  {\widetilde \psi}_{20}(\chi)&=& S_{20}(\chi) +bS_{21}(\chi) +
  \frac{(2b)^2}{3^2}S_{22}(\chi).
\end{eqnarray}
In the notations of (\ref{basis_change}),
\begin{eqnarray}
{\widetilde {\boldsymbol {\psi}}}_{1}(\chi)=\left(\begin{array}{c}
{\widetilde \psi}_{11}(\chi)\\
{\widetilde \psi}_{10}(\chi)
\end{array}
\right)={\mathbf A}^{(1)}{\mathbf S}_{1}(\chi),
&&
{\boldsymbol{S}}_{1}(\chi)=
\left(
\begin{array}{c}
  S_{11}(\chi)\\
    S_{10}(\chi)
  \end{array}
\right),\nonumber\\
{\mathbf A}^{(1)} &=&
\left(
\begin{array}{cc}
  1&0\\
  b&1
  \end{array}
\right), 
\label{so21_2}
\end{eqnarray}

and

\begin{eqnarray}
  {\widetilde {\boldsymbol{\psi}}}_{2}(\chi)=\left(
  \begin{array}{c}
    {\widetilde \psi}_{22}(\chi)\\
    {\widetilde \psi}_{21}(\chi)\\
{\widetilde \psi}_{20}(\chi)
\end{array}
  \right)={\mathbf A}^{(2)}{\mathbf S}_2(\chi),&\quad & {\mathbf S}_{2}(\chi)=\left( \begin{array}{c}
    S_{22}(\chi)\\
    S_{21}(\chi)\\
    S_{20}(\chi)\end{array}\right),\nonumber\\
 {\mathbf A}^{(2)}&=& \left(
  \begin{array}{ccc}
1&0&0\\
    \frac{2b}{3}&1&0\\
    \frac{(2b)^2}{3^2}&b&1
    \end{array}
  \right),
\label{so121_3}
\end{eqnarray}
it is easily checked that,
\begin{eqnarray}
  {\mathcal A}^{(1)\to ( 2) }_+ {\widetilde {\boldsymbol {\psi}}}_{1}(\chi)
    ={\widetilde {\boldsymbol{\psi}}}_{2}(\chi),&&\nonumber\\
{\mathcal A}^{(1)\to (2)}_+ =  {\mathbf A}^{(2)} \left(
  \begin{array}{cc}
    \sin\chi&0\\(\sin\chi)^{-1} A_+\sin\chi &0\\
    0&(\sin\chi)^{-1} A_+\sin\chi
    \end{array}\right)
  \left[{\mathbf A}^{(1)}\right]^{-1},&&
    \label{so21_4}
\end{eqnarray}
with $A_+$ in (\ref{Apm}), hold true and thereby, with the aid of the (\ref{so12_1}),(\ref{so121_3}) it is possible
to recognize in ${\mathcal A}^{(1)\to (2)}_+$ an intertwining relation of the type in (\ref{Al2}). Through the  similarity transformation by $\sin\chi$ applied to the $A_+$ operator, we accounted for (\ref{Gl2}) and hence for the fact that we are considering the algebra on the hyper-sphere, while $A_\pm$ have been introduced at the level of the 
one-dimensional Schr\"odinger equation in (\ref{Gl1}).

The rectangular $3\times 2$ matrix in (\ref{so21_4}) plays in the subspace of the ${\widetilde {\boldsymbol{\psi}} }_K(\chi)$ functions the role of a ladder operator conserving the $\ell$ label and rising the $K$ quantum number by one unit. In a same way the lowering ladder operator can be constructed. The ${\mathcal A}^{(K)\to (K)}_3$ operator reads as the usual $A_3$, multiplied by the unit matrix, and the
intertwining happens by  one and the same quadratic matrix of dimensionality $(K+1)\times (K+1)$.

Back to (\ref{so12_1}), the  transition between the states  of our interest is now directly read off from (\ref{so21_4})
as,
\begin{equation}
  F_{21}(\chi)=e^{\frac{\alpha_2\chi}{2}} \left(\frac{2b}{3}\sin\chi +
  (\sin\chi)^{-1} A_+ \sin\chi \right)
  e^{-\frac{\alpha_1\chi}{2}}F_{11}(\chi).
\label{end_answ}
\end{equation}
Along these lines all possible ladder operators can be constructed.
Such operators are of interest in spectroscopic studies, where they can be used
in the description of  excitation- and de-excitation modes.

As long as intertwining transformations conserve the commutators, the $so(2,1)$ algebra constituted by $A_\pm$ and $A_3$ from the respective equations
(\ref{Apm}), and (\ref{a3}), remains same also in the representation induced by the metric deformation under consideration. 
 
Admittedly, the operators constructed in this way are representation dependent, while in  canonical  Lie group algebras they are representation independent. A similar peculiarity is also observed in other  cases of designing  algebras of dynamical symmetries, among them the $so(2,4)$ algebra of  dynamical
conformal symmetry for the H Atom, a subject still under clarification and on which one finds relatively few assertive discussions in the literature (see \cite{Levai} and references therein).

In the literature,  attempts have  been undertaken \cite{Levai}, \cite{Koc}  to construct representation independent $so(2,1)$ ladder operators directly at the level of the polynomial parts of the wave functions in (\ref{sol_CCl}). To the amount the parameters of the orthogonal polynomials defining the wave functions in the problem under consideration  are $K$ dependent, the ladder operators have to rise/lower  this number by one unit in the parameters too, i.e. they have to secure,
\begin{equation}
\beta_K\to \beta _{K+1}, \quad \alpha_K\to \alpha_{K+1}.
\label{parm_lddr}
\end{equation}
However,  recurrence formulas ensuring such a change in the parameters of the
$R_n^{\alpha_K,\beta_K}(\cot \chi)$ polynomials exist 
only for the $\beta_K$ parameter that is linear in $K$, while  for the parameter $\alpha_K$, containing the inverse of $(K+1)$ (c.f. (\ref{sol_CCl}))  they can not be constructed. For this reason, in the literature, the $so(2,1)$
symmetry at the end could not  be realized in a representation independent fashion.  Rather, the magnitude  of the cotangent potential has been altered toward the representation dependent magnitude of, 
 $(\pm 2b(K+1) )$, while  the original universal  $(\pm 2b)$ value was given up. In so doing, the contribution of the  Balmer-like  term  in the energy dies out and is reduced to a mere universal constant.

We have circumvented the above difficulty in a twofold way, namely by first admitting, similarly to the Hydrogen atom case, representation-dependent operators, like the one in (\ref{end_answ}),  and secondly, through the incorporation of the Balmer-like term, viewed as  the  representation dependent constant, $\frac{\alpha_K^2}{4}$, from  (\ref{sol_CCl}), into the potential (\ref{Hrm_PT}).
Remarkably, through the conformal deformation of the metric of the compactified Minkowski space, which we  worked out in the above Sect.~8.2, precisely this interaction, termed to as  conformal charge-dipole, and a version of the well known trigonometric Rosen-Morse potential, emerged  in (\ref{Hrm_PT}).
Afterward, the Balmer-like term can always been passed to the side of the eigenvalues and get absorbed by the energy, if needed.

As a result, the conformal symmetry of the conformal charge dipole potential in (\ref{Hrm_PT})  has been made ma\-ni\-fest representation by representation, much alike to the case of the dynamical $so(2,4)$ symmetry of the  Hydrogen atom.
Be\-cau\-se in add\-ition the equation (\ref{Frnds3})  is  also  most adequate for  the description of dipole systems with con\-fined char\-ges, such as the  color-elec\-tric char\-ges in quark-anti\-quark systems , the mesons, it shows itself as an option for addressing QCD from quantum mechanics.

\acknowledgments
This work has been partially supported by the Bulgarian National Science Fund research grant
DN 18/3.

\end{document}